\documentclass[11pt]{article}
\usepackage{jheppubmodnew}
\pdfoutput=1
\usepackage{hyperref}
\usepackage{psfrag}
\usepackage{array}
\usepackage{amssymb}
\usepackage{amsmath}
\usepackage{amsthm}
\usepackage{graphicx}
\usepackage{etoolbox}
	
\usepackage[punctsep]{collref}
\collectsep[]{;}
\usepackage{bm}
\usepackage{epsfig}
\usepackage{wrapfig}
\usepackage{tikz}
\usepackage{mathtools}
\usepackage{tensor}
\usepackage{cleveref}
\usetikzlibrary{arrows,plotmarks,positioning,calc}

\def\pb[#1,#2]{\{#1, #2\}}
\def\deb[#1,#2]{[#1,#2]_{\text{D.B.}}}

\def\l{\lambda}

\def\w{\omega}

\def\a{\alpha}

\def\Or[#1]{\text{O}\!\left(#1\right)}
\def\dotl[#1,#2]{\left\langle #1,\, #2 \right\rangle}
\def\dotlb[#1,#2]{\left\langle #1,\, #2 \right\rangle}
\def\dotlm[#1,#2]{\left[ #1,\, #2 \right]}
\def\dotp[#1,#2]{(\vect{#1} \cdot\vect{#2})}
\def\aff[#1,#2]{\hat{#1}(#2)}

\def\n4sym{{\cal N}=4 SYM}
\def\>{\rangle}
\def\<{\langle}

\def\projsho[#1]{{\cal P}^{\text{sho}}_{#1}}
\def\transsho[#1,#2]{{\cal T}^{\text{sho}}_{#1,#2}}
\def\weight[#1,#2,#3]{\{(#1),#2,#3\}}
\def\ads[#1]{$\text{AdS}_{#1}$}

\newcommand{\be}{\begin{equation}}
\newcommand{\ee}{\end{equation}}
\newcommand{\ba}{\begin{align}}
\newcommand{\ea}{\end{align}}
\newcommand{\bs}{\begin{split}}
\def\sess\end{split}
\newcommand{\vect}[1]{{\boldsymbol{#1}}}

\newcommand{\bea}{\begin{eqnarray}}
\newcommand{\eea}{\end{eqnarray}}

\def \bes {\begin{equation*}}
\def \ees {\end{equation*}}

\def \b  {\beta}

\def\alcut[#1]{{\cal A}_{#1, \epsilon}}
\def\alseg[#1,#2]{{\cal B}_{#1, #2}}

\def\supcharge[#1]{\{#1\}}

\def\projsupeig[#1]{{\cal P}_{{\ell, m}}[{#1}]}
\def\transop[#1, #2]{T_{\{#1\}, \{#2\}}}
\def\supket[#1]{|\{#1\} \rangle}
\def\supbra[#1]{\langle \{#1\} | }

\def\rsop[#1]{X_{#1}}

\def\projlow[#1,#2]{P_{{#1}<{#2}}}
\def\rd[#1]{^{(#1)}R}

\def\gser{X}
\def\matser{Y}

\def\solfunc{{\cal F}}

\def\qstate{\psi}
\def\sodminone{{\mathrm{SO}}(d-1)}
\def\confgrp{{\mathrm{SO}}(1,d+1)}

\def\doubleder{{\cal D}_{\cal F}}

\def\conf{\Omega}

\def\coeff[#1,#2]{{\mathcal G}_{#1,#2}}
\def\dcoeff[#1,#2]{\delta {\mathcal G}_{#1,#2}}
\def\tcoeff[#1,#2]{{\widetilde{\mathcal G}}_{#1,#2}}
\def\lcoeff[#1,#2]{{\mathcal G}^{\lambda}_{#1,#2}}
\def\dgtr{\delta_g}

\def\diff{{\delta}_{\xi,\vphi}}
\newcommand\diffWeyl{diff $\times$ Weyl}
\def\vI{\vec{i}}
\def\vJ{\vec{j}}

\def\innerp[#1,#2]{\left({#1, #2} \right)}

\usepackage{cite}

\def\pg{\paragraph}

\def\eg{\emph{e.g.} }
\def\ie{\emph{i.e.} }

\def\nt{\notag}

\def\cA{\mathcal{A}}

\def\cF{\mathcal{F}}
\def\cG{\mathcal{G}}
\def\cH{\mathcal{H}}

\def\cL{\mathcal{L}}

\def\cO{\mathcal{O}}

\def\cS{\mathcal{S}}

\def\p{\partial}

\def\/{\over}

\def\rn{\rangle}
\def\ln{\langle}

\def\vphi{\varphi}
\def\a{\alpha}
\def\b{\beta}
\def\d{\delta}
\def\k{\kappa}
\def\g {\gamma}

\def\w {\omega}

\def\l{\ell}

\def\n {\nabla}
\def\L{\Lambda}
\def\D{\Delta}

\def\Om {\Omega}

\def\ra{\rightarrow}

\def\r{\mathrm}

\def\_{\hspace{2cm}}
\def\-{\\\notag}
\def\={&=&}

\newcommand{\bpm}{\begin{pmatrix}}
\newcommand{\epm}{\end{pmatrix}}

\newcommand{\bit}{\begin{itemize}}
\newcommand{\eit}{\end{itemize}}

\newcommand{\ben}{\begin{enumerate}}
\newcommand{\een}{\end{enumerate}}

\newcommand\bsp{\begin{split}}
\newcommand\esp{\end{split}}

\def\le{\left}
\def\ri{\right}

\def\l{\ell}

\def\qq{\qquad}

\newcommand{\enc}[1]{\left( #1\right)}
\newcommand{\encsq}[1]{\left[ #1\right]}

\def\qstate{\psi}

\newcommand{\changelocaltocdepth}[1]{%
  \addtocontents{toc}{\protect\setcounter{tocdepth}{#1}}%
  \setcounter{tocdepth}{#1}%
}

\title{The Hilbert space of de Sitter quantum gravity}
\author[a]{Tuneer Chakraborty,}
\author[a,b]{Joydeep Chakravarty,}
\author[a]{Victor Godet,}
\author[a]{Priyadarshi Paul}
\author[a]{and Suvrat Raju}

\affiliation[a]{International Centre for Theoretical Sciences (ICTS-TIFR),\\ Tata Institute of Fundamental Research,
Shivakote, Hesaraghatta, Bengaluru 560089, India}

\affiliation[b]{Department of Physics, McGill University,\\
3600 Rue University, Montreal, H3A 2T8, QC Canada}

\emailAdd{tuneer.chakraborty@icts.res.in}
\emailAdd{joydeep.chakravarty@mail.mcgill.ca}
\emailAdd{victor.godet@icts.res.in}
\emailAdd{priyadarshi.paul@icts.res.in}
\emailAdd{suvrat@icts.res.in}
\date{}

\abstract{We obtain solutions of the Wheeler-DeWitt equation with positive cosmological constant for a closed universe  in the large-volume limit. We argue that this space of solutions provides a complete basis for the Hilbert space of quantum gravity in an asymptotically de Sitter spacetime. Our solutions take the form of a universal phase factor multiplied by distinct  diffeomorphism invariant functionals, with simple Weyl transformation properties, that obey the same Ward identities as a CFT partition function.  The Euclidean vacuum corresponds to a specific choice of such a functional but other choices are equally valid. Each functional can be thought of as specifying a ``theory'' and, in this sense, the space of solutions is like ``theory space''.  We describe another basis for the Hilbert space where all states are represented as excitations of the vacuum that have a specific constrained structure. This gives the finite $G_N$ generalization of the basis proposed by Higuchi in terms of group averaging, which we recover in the nongravitational limit. 
}

\begin{document}
\maketitle
\section{Introduction}
In this paper, we seek to address a basic question about quantum gravity in asymptotically de Sitter space: what is the space of states in such a theory? 

This question has received surprisingly little attention. Considerable attention has been devoted to the Hartle-Hawking state, or the Euclidean vacuum, which is obtained by performing the path integral on a Euclidean space with only one boundary \cite{Hartle:1983ai} and can be obtained by analytic continuation from AdS \cite{Maldacena:2002vr}. It is sometimes erroneously believed that other states in the Hilbert space can be obtained, as in a nongravitational quantum field theory, by simply acting with arbitrary field operators on the vacuum.

Higuchi \cite{Higuchi:1991tk,Higuchi:1991tm} pointed out that a naive Fock-space construction does not lead to the correct Hilbert space, even for weakly-coupled gravity. Even as the gravitational coupling is taken to zero, it is necessary to impose the constraints of the gravitational Gauss law on the Fock space.  Since the Cauchy slices in de Sitter space are compact, the Gauss law implies that states must have zero charges under the de Sitter isometries  \cite{1973CMaPh..32..291B,1975JMP....16..493M,1976JMP....17.1893M}.  At first sight, it would appear that this constraint excludes all states except for the Euclidean vacuum. 

Higuchi proposed an ingenious construction, where one starts with a ``seed state'' and then averages it over the de Sitter-isometry group so as to produce invariant states. These states are not normalizable in the original norm, but Higuchi also proposed a modified norm, which amounts to dividing the QFT norm of these states by the infinite volume of the de Sitter-isometry group. It was later checked, that in some examples, the above prescription leads to a well-defined norm \cite{Marolf:2008hg,Marolf:2008it}. 

In this paper, we will systematically investigate the form of the Hilbert space by studying solutions to the Wheeler-DeWitt (WDW) equation \cite{DeWitt:1967yk} in a theory of gravity coupled to matter.  The WDW equation always constrains the set of allowed states in any theory of gravity, but it is often impossible to solve it and obtain the structure of the Hilbert space.  However, here we show that the equation simplifies in the limit where the cosmological constant dominates over the intrinsic curvature of the Cauchy slices and other terms in the local ``energy density''.  We will study the case where this condition holds at every point on the Cauchy slice and we refer to this as the ``large-volume limit''.  This large-volume limit is different from the perturbative limit that we studied in \cite{Chowdhury:2021nxw}, and  we will argue in section \ref{secsol} that the key structural properties of the solutions we find are valid at all orders in perturbation theory.

Physically, this limit is very easy to understand. An asymptotically de Sitter spacetime attains the large-volume limit at asymptotically late times or early times. Therefore our solution to the WDW equation can be thought of as a form of ``asymptotic quantization'' --- a program  \cite{Ashtekar:1981sf,Ashtekar:1981hw,Ashtekar:1987tt} that can be applied to the full nonlinear theory and has been fruitful in understanding the structure of the Hilbert space in asymptotically flat spacetimes. 

We will show below that the solutions to the WDW equation in this limit can be characterized by diffeomorphism invariant wavefunctionals that have a simple specified behaviour under Weyl transformations. The Euclidean vacuum is also described by such a wavefunctional and is known to have these properties. But the new result in this paper is that {\em all} valid wavefunctionals have the same simple behaviour under diffeomorphisms and Weyl transformations. 

These wavefunctionals can be expanded in terms of the fluctuations of the metric and other degrees of freedom. The coefficient functions that appear in this expansion obey the same Ward identities as CFT correlators. So, one way to understand our result is that the space of solutions to the WDW equation in asymptotically-de Sitter space is described as ``theory space'' \ie if one is given a set of correlation functions that obey the Ward identities imposed by conformal invariance (but not necessarily the constraints of unitarity or locality)  then they can be used to construct a solution to the constraints.

Moreover, we will show that Higuchi's prescription for group averaging emerges naturally as the weak-coupling limit of such solutions. Therefore our analysis validates Higuchi's ansatz in the limit of weak coupling but also explains how it must be generalized in the interacting theory. 

While we provide a complete basis for the vector space of allowed states in the theory in this paper, it is necessary to define a norm on this space to complete the Hilbert-space structure.  We define the norm in a companion paper \cite{dssecond2023}. There, we also discuss the meaning of cosmological correlators when quantum-gravity effects are taken into account, and establish the principle of holography of information in asymptotically de Sitter space.

\paragraph{\bf Relation to previous work.} Our results are entirely consistent with the observation that the wavefunctional of the Euclidean vacuum  can be computed in terms of the partition function of an appropriate CFT after dressing it with an appropriate phase factor \cite{Strominger:2001pn,Maldacena:2002vr,Hertog:2011ky,Anninos:2012ft,Castro:2012gc,McFadden:2009fg,Halliwell:2018ejl}. This is a useful observation. However, the Euclidean vacuum is a single wavefunctional.\footnote{As we discuss in \cite{dssecond2023}, naively, this state does not appear to be normalizable therefore it might not even be part of the Hilbert space.} It does satisfy the WDW equation but it is not the unique wavefunctional that does so \cite{Banks:1984cw}. Our objective in this paper is to systematically consider the space of all solutions to the constraints. 

An interesting proposal for the Hilbert space was made from a top-down perspective in \cite{Anninos:2017eib} (building on \cite{Anninos:2011ui}) for a specific theory with a low-energy description as Vasiliev gravity \cite{Vasiliev:2003ev}. Our approach is complementary since it is ``bottom up'' and starts from the bulk.  Since the answer in \cite{Anninos:2017eib} is provided in terms of an auxiliary set of scalars on the late-time boundary, we cannot immediately compare our proposed answer with \cite{Anninos:2017eib} but it is an interesting open problem to perform this comparison.

The WDW equation was also recently studied in AdS \cite{Araujo-Regado:2022gvw, Witten:2022xxp}.  It would be interesting to apply these techniques to dS. See \cite{Halliwell:2018ejl,Hertog:2011ky,Hartle:2008ng} for earlier analyses of the constraints,  \cite{Araujo-Regado:2022jpj,Blacker:2023oan} for recent progress in this direction and \cite{Park:2015qxa} for related discussion.

There have also been suggestions \cite{fischler:talk,Banks:2000fe,Witten:2001kn,Banks:2001yp,Arenas-Henriquez:2022pyh,Arias:2019pzy,Banks:2018ypk,Banks:2020zcr} that the Hilbert space in de Sitter space should be finite dimensional. This might happen due to nonperturbative effects that constrain the allowed form of states, but our analysis does not shed light on this issue.

\section{Summary of results}

\begin{figure}
	\centering
	\includegraphics[width=7cm]{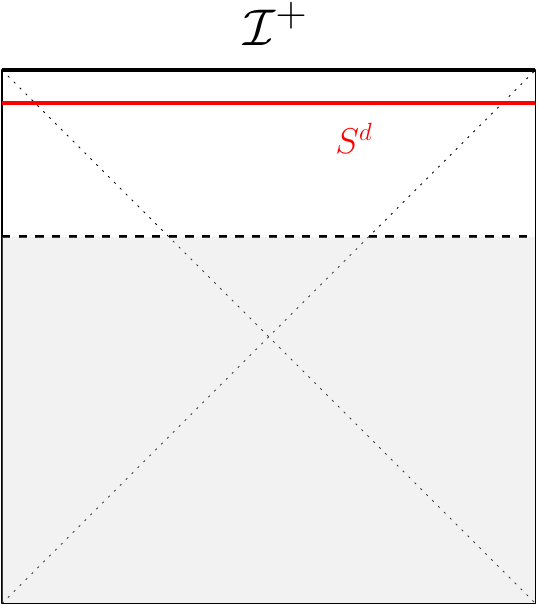}
	\caption{\em We are considering a late time slice (in red) with topology $S^d$ in an asymptotically de Sitter spacetime. In the late time expansion, the Wheeler-DeWitt equation can be solved and, up to a universal phase factor, the space of solutions  is  the space of functionals that transform under \diffWeyl{} in the same way as CFT$_d$ partition functions.}\label{fig:penrose}
\end{figure}

We study the WDW equation (reviewed below) in the regime where the cosmological constant dominates pointwise over the Ricci scalar and over the matter potential.  
 In this limit the Cauchy slices, which are topologically $S^{d}$, grow to a very large volume. This can be thought of as a late time slice in an asymptotically dS$_{d+1}$ spacetime, see Figure \ref{fig:penrose}.

 It is convenient to work in a coordinate system where this limit corresponds to a large conformal factor $\conf(x)=\text{det}(g)^{1/2d}$ for the metric. In section \ref{secsol}, we find the solutions to the WDW equation take the following form when expressed as functionals of the metric fluctuations and matter fields.
\be
\label{psiasympform}
\Psi[g, \chi] \underset{\conf \rightarrow \infty}{\longrightarrow}  e^{i S[g, \chi]}Z[g,\chi] ~,
\ee
where $S[g, \chi]$ is a universal phase factor obtained by integrating local densities.  $Z[g,\chi]$ is, in general, a nonlocal functional of $g$ and $\chi$ that is diffeomorphism invariant and transforms in a simple way under Weyl transformations 
\be
\conf {\delta Z[g, \chi] \over \delta \conf (x)} = {\cal A}_{d}[g] Z[g,\chi] ~,
\ee
where the anomaly polynomial ${\cal A}_{d}[g]$ can be computed explicitly. The anomaly polynomial vanishes for $d$ odd and is imaginary for $d$ even  and therefore $|Z[g, \chi]|^2$ is a \diffWeyl{} invariant functional.  The explicit form of $\cA_d$ depends on a choice of normal ordering and may be corrected at higher orders in $\kappa$. However, we argue that the structural form \eqref{psiasympform} is valid at all orders in perturbation theory.

At the cost of a possible phase, we can perform a Weyl transformation to study $Z[g,\chi]$ in the vicinity of the flat metric $g_{i j} = \delta_{i j} + \kappa h_{i j}$. Then, provided $\log\,Z[g,\chi]$ is well behaved in the limit $g_{i j} \to \delta_{i j}, \chi \to 0$, we can expand it as 
\be
\log Z[g,\chi] = \sum_{n,m} \kappa^n \coeff[n,m] ~,
\ee
where $\coeff[n,m]$ are multilinear functionals of the metric fluctuation, $h_{i j}$ and matter fluctuations of order $n$ and $m$ respectively. In section \ref{sechilbastheory} we derive a set of Ward identities that constrain the form of $\coeff[n,m]$. If one writes
\be
\label{gnmdef}
\coeff[n,m] = {1 \over n! m!} \int d \vec{y} d \vec{z} \,G^{\vI\vJ}_{n,m}(\vec{y}, \vec{z}) h_{i_1 j_1}(z_1) \ldots h_{i_n j_n}(z_n) \chi(y_1) \ldots \chi(y_m)  ~,
\ee
then the coefficient functions $G^{\vI\vJ}_{n,m}$ obey the same Ward identities as those obeyed by a connected correlator of $n$ stress tensors and $m$ operators of dimension $d - \Delta$, where $\Delta$ is related to the mass of the scalar field by \eqref{massFormula}. 

The Euclidean vacuum is a particular state of the form above. Our new result is that {\em all} solutions are spanned by functionals of this form and related in a simple manner to \diffWeyl{} invariant functionals.

Since these identities relate functions with different values of $n$, the different coefficient functions $G^{\vec{i} \vec{j}}_{n,m}$ are not independent of each other. A complete set of such correlation functions satisfying a set of mutually consistent identities can be said to define a ``theory''. Therefore our space of solutions can be thought of as ``theory space''. Of course, we emphasize that this theory is not a unitary, or even a local, CFT.

 If we represent the Euclidean vacuum  by $e^{i S[g,\chi]} Z_{0}[g, \chi]$ then we show in section \ref{secpert} that a convenient basis for the space of states is given by wavefunctionals of the form
\be
\label{summaryseries}
\Psi[g, \chi] = e^{i S[g, \chi]}\sum_{n,m}  \kappa^n \dcoeff[n,m] Z_0[g,\chi] ~,
\ee
where $\dcoeff[n,m]$ is the {\em difference} of two sets of multilinear functionals, each of which is of the form \eqref{gnmdef}.

The differences $\dcoeff[n,m]$ also obey Ward identities that relate $\dcoeff[n+1,m]$ to $\dcoeff[n,m]$ and therefore the series in \eqref{summaryseries} is infinite. However, in the limit $\kappa \rightarrow 0$, it is possible to focus on a single term in \eqref{summaryseries}.  We show that the nongravitational states so obtained correspond {\em precisely} to the group-averaged states found by Higuchi. For each state, we also explicitly find the corresponding ``seed state''.  

Therefore our analysis justifies Higuchi's proposal in the nongravitational limit. However, it also reveals how Higuchi's prescription must be generalized away from zero coupling. At nonzero $\kappa$, one must add the terms required by the Ward identities to complete the series \eqref{summaryseries}.

\section{Asymptotic solutions to the Wheeler-DeWitt equation \label{secsol}}
In this section, we will show that  solutions to the Wheeler-DeWitt equation take on the asymptotic form displayed in \eqref{psiasympform}.

We consider Einstein gravity in $d+1$ dimensions with a positive cosmological constant in units where
\be\label{ccvalue}
\Lambda = {d (d-1) \over 2 }~.
\ee
In a $d+1$ split, the spacetime metric can be written in the form  
\be
ds^2 = -N^2 dt^2 + g_{i j} (d x^i + N^i d t)(d x^j + N^j d t) ~,
\ee
where $N$ and $N^i$ are the shift and lapse functions \cite{Arnowitt:1962hi,misner1973gravitation}.
The spatial slices are taken to be compact with metric $g_{ij}$. In addition, we might have matter degrees of freedom in the theory. As an illustration, we consider a massive scalar $\chi$ although we do not expect that our results will depend on the choice of matter.

In the canonical formalism, a state in such a theory is represented by a wavefunctional $\Psi[g_{i j}, \chi]$ that assigns an amplitude to a particular configuration of the metric and the matter fields on a spatial slice.  This wavefunctional must obey the Hamiltonian and momentum constraints that arise simply by imposing diffeomorphism invariance on the theory \cite{DeWitt:1967yk}
\be
\label{hpconstraint}
{\cal H}\, \Psi[g_{i j}, \chi] = 0~, \qquad {\cal H}_{i} \,\Psi[g_{i j}, \chi] = 0~.
\ee
The Hamiltonian constraint is
\be
\label{hamexplicit}
\cH =2 \k^2 g^{-1}\le(  g_{i k} g_{j l} \pi^{kl }\pi^{ij} -{1\/d-1}(g_{ij}\pi^{ij})^2\ri)  - {1\/2\k^2}(R-2\L) + \cH_\r{matter} + \cH_{\text{int}} ~,
\ee
and the equation in \eqref{hpconstraint} setting it to annihilate the wavefunctional is called the Wheeler-DeWitt (WDW) equation. The momentum constraint is
\be
\cH_i = -2 g_{ij} \n_k {\pi^{jk} \over \sqrt{g}} +\cH_{i,\r{matter}}~.
\ee
The gravitational coupling is $\kappa^2 = 8\pi G_N$. The momentum operator acts on the wavefunctional as
\be
\pi^{i j} = -i{\d \over \d g_{i j}}~.
\ee
We take the matter energy density to be of the form
\be
\cH_{\r{matter}} = {1 \over 2}g^{-1} \pi_{\chi}^2  + V_{\text{matter}}; \qquad V_{\text{matter}} = {1 \over 2} g^{i j} \partial_{i} \chi \partial_{j} \chi + {1 \over 2} m^2 \chi^2 ~,
\ee
and $\cH_{i,\r{matter}} = {1\/\sqrt{g}}\pi_\chi \p_i\chi$ is the matter momentum density.

The self-interactions of matter, its interaction with gravity and also potentially higher-order interactions have all been included in  $\cH_{\text{int}}$. The analysis that follows will be largely insensitive to the details of $\cH_{\text{int}}$.

\subsection{Asymptotic expansion}

The Hamiltonian constraint \eqref{hamexplicit} has terms that involve functional derivatives, which we can call ``kinetic terms'', and terms without functional derivatives that we can call ``potential terms''.  Here, we will study the equation in the regime where the cosmological constant dominates over all other potential terms everywhere on the Cauchy slice. In particular, this means that the Ricci scalar and the matter potential are very small compared to the cosmological constant,
\be
\label{assumlambdadom}
R \ll \Lambda; \qquad V_{\text{matter}} \ll \Lambda~.
\ee
Classically, we expect from the cosmic no hair theorems that when \eqref{assumlambdadom} is met, the geometry will be that of de Sitter space (see \cite{Anninos:2012qw} for a nice discussion and references). We will find that the WDW equation also simplifies. Since these conditions must apply everywhere on the spatial slice, our analysis does not apply to geometries that have singularities on the Cauchy slice under consideration.

\paragraph{\bf An intrinsic notion of time.}   We will present solutions to \eqref{hpconstraint}, which are valid when the assumption \eqref{assumlambdadom} is met. Our physical interpretation is that these solutions describe ``late times'' in an asymptotically de Sitter universe. This includes states that might have very complicated features at finite times but settle down asymptotically to de Sitter space.  

However, the equations \eqref{hpconstraint} do not make any reference to time. Nor, in the case of de Sitter space, do we have an asymptotic boundary that can be used to set up an external clock.  It was pointed out long ago by DeWitt \cite{DeWitt:1967yk}, that this problem can be addressed by using an intrinsic observable as a clock. Correlators of other observables with this intrinsic clock then provide a notion of how the state varies with ``time''. 

Here, we note that \eqref{assumlambdadom} also implies that the volume of the Cauchy slice 
\[
\log\,\int  d^d x\,\sqrt{g}
\]
becomes very large. So we can use the logarithm of the volume, which is a dimensionless quantity in the units chosen above, as a clock. This notion of time is related to the so-called ``York time''\cite{York:1972sj}. This provides an operational meaning to the phrase ``late time''.   

If one studies a spacetime that contracts from an infinitely large volume in the asymptotic past, then our analysis also applies to asymptotically early times when \eqref{assumlambdadom} is met.  But to ask questions about ``finite times'', one must necessarily go beyond the assumption \eqref{assumlambdadom} somewhere on the slice. We will not address this regime here.

\paragraph{\bf Intermediate variables.} To facilitate our analysis, we will introduce  {\em intermediate variables}, $\conf$ and $\g_{ij}$ and write the metric on the spatial slice as
\be
\label{rescalemet}
g_{ij}  = \conf^2 \g_{ij} ~,
\ee
where $\det(\g_{i j}) = 1$. In terms of the original degrees of freedom, we define
\be
\label{gtoconfgamma}
\conf = (\det(g_{i j}))^{1 \over 2 d}; \qquad \g_{i j} = g_{i j} \det(g_{i j})^{-{1 \over d}}~.
\ee
Subject to the assumption \eqref{assumlambdadom}, it should be possible to find  a coordinate system where $\conf$ is everywhere large and we will assume that such coordinates have been chosen.

It is also expected on physical grounds that the density of matter fields  will get ``diluted'' as the scale factor increases. This leads us to define a set of intermediate variables $O$ for the matter fields according to
\be
\label{rescalechi}
\chi  = \conf^{-\Delta} O~.
\ee
In the analysis below, we will fix $\D$ in terms of the mass of the field and the cosmological scale. 

We emphasize that \eqref{rescalemet} and \eqref{rescalechi} correspond to an exact change of variables, and so the content of the equations \eqref{hpconstraint} is preserved. Second, we note that the split of the original metric into a Weyl factor and a Weyl-invariant part is coordinate dependent. Nonetheless, we will be careful to write all our final answers in a diffeomorphism-invariant form in terms of the original variables $g$ and $\chi$. So the reader should think of the change of variables simply as an intermediate technical trick.

\subsection{Solution algorithm \label{subsecsolalg}}

In this subsection we outline and implement an algorithm to find solutions to the constraints in the limit \eqref{assumlambdadom}. This subsection is somewhat technical and the reader who is interested just in the results can jump ahead to subsection \eqref{subsecasympsol}.

Our procedure to find a solution to the constraints has three steps.
\begin{enumerate}
\item
We rewrite the Hamiltonian constraint in terms of the conformal variables, $\conf$ and $\g$. 
\item
We then seek a solution where the wavefunctional can be represented as the exponential of a functional, $\solfunc$, that can be expanded in a series of terms  that have a distinct scaling at large $\conf$. With suitable assumptions about normal ordering, the Hamiltonian constraint can be written in terms of $\solfunc$.
\item
The functional $\solfunc$ involves two distinct series expansions --- one that is present even for pure gravity and another that involves the matter fields. We first solve for the gravitational part and then for the matter part.
\end{enumerate}
The solution that we present below can be thought of as an elaboration of the solution to the radial WDW equation given in AdS by Freidel \cite{Freidel:2008sh}.

\paragraph{\bf Rewriting the constraints.}
It is shown in Appendix \ref{subapprewriteconst} that, in terms of these new variables, the Hamiltonian constraint can be rewritten as
\be
\begin{split}\label{hrewritten}
&\cH = {2\k^2 \over \conf^{2 d}} \le[ {1\/4d(d-1)} \le(\conf {\delta \over \delta \conf}  + \D  O {\delta \over \delta O}  \ri)^2- \le(\gamma_{ik} \gamma_{j\l} - {1 \over d} \gamma_{ij} \gamma_{k\l} \ri) {\delta \over \delta \gamma_{ij}} {\delta \over \delta \gamma_{k\l}} \ri] \\ &+ {\L\/\k^2} -{1\/2\k^2}R[\conf^2\gamma]
 -{1\/2} \conf^{2 (\Delta-d)} {\delta^2 \over \delta O^2}+{1\/2} m^2 \conf^{-2 \D} O ^2+{1\/2}\conf^{2} \g^{ij}\p_i (\conf^{-\Delta} O) \p_j (\conf^{-\Delta} O) + \cH_{\text{int}}~.
\end{split}
\ee
We do {\em not} seek to rewrite the momentum constraint. This is because the momentum constraint simply imposes that the wavefunctional is invariant under $d$-dimensional diffeomorphisms. This can be seen more easily in terms of the original variables $g$ and $\chi$ that transform as tensors rather than tensor densities.

\paragraph{\bf Series expansion.}
We expand the solution to \eqref{hpconstraint} as 
\be
\label{solansatz}
\Psi = e^{i \solfunc}; \qquad \solfunc = \sum_{n=0}^\a  \gser_{\alpha - n}  + \sum_{m=0}^{m_\b} \matser_{\beta - m} + \Or[{1 \over \conf}]~.
\ee
Here, the functionals $\gser_k$ and $\matser_k$ are undetermined functionals that grow as $\conf^k$ at large $\conf$.   More precisely we have
\be
\label{asympscaling}
\gser_k, \matser_k\sim \Om^k, \qq \Om\to+\infty ~,
\ee
except for $k=0$ in even $d$ where we find an anomalous term that can be thought of as scaling with $\log(\conf)$. For now, we keep $\alpha$ and  $\beta$ as undetermined parameters. It will turn out below that the solution to the WDW equation will require a series that grows with integer powers of $\conf$, corresponding to $\alpha = d$, and a series that grows with non-integer powers, $\beta = d - 2 \Delta$. The maximum value of $m$ is $m_\b =\lfloor\r{Re}\,\b\rfloor$.  

It is justified to separate the two series since $\beta$ can be varied by varying parameters in the theory, and therefore, the two kinds of terms can be distinguished at a generic point in parameter space.  The entire expression \eqref{solansatz} has an undetermined remainder denoted by $\Or[{1 \over \conf}]$ which corresponds to terms that decay at large $\conf$. We will not work out the specific form of this remainder in this paper. It is the term that is necessary to understand ``finite-time'' physics.

We do not assume that the functionals $\gser_{\alpha-n}$ and $\matser_{\beta - m}$ are local functionals of $g_{i j}$ and $\chi$. It will turn out however
that  the leading terms in the series $\gser_{\alpha-n}$ will depend on local functionals of the metric. These correspond to the ``gravitational part'' of the solution.  However, $\gser_0$ will be, in general, a nonlocal functional that  depends both on $g_{i j}$ and $\chi$. The series $\matser_{\beta - m}$ corresponds to the ``matter part'' of the solution and will comprise local functions of $g_{i j}$ and $\chi$.

\paragraph{\bf Normal ordering and simplification.} 
Acting with the rewritten constraint \eqref{hrewritten} on the ansatz \eqref{solansatz} we find that the WDW equation can be written as
\be
\label{Hpsi}
\begin{split}
 0 &= -2\k^2 {1 \over \conf^{2 d}} \le[ {1\/4d(d-1)} \le(\conf {\delta \solfunc \over \delta \conf}  + \D O {\delta \solfunc \over \delta O}  \ri)^2- \le(\gamma_{ik} \gamma_{j\l} - {1 \over d} \gamma_{ij} \gamma_{k\l} \ri) {\delta \solfunc \over \delta \gamma_{ij}} {\delta \solfunc \over \delta \gamma_{k\l}} \ri] + \doubleder \\ &
  + {\L\/\k^2} -{1\/2\k^2}R[g]+ {1\/2}  \conf^{2 (\Delta-d)} \le({\delta \solfunc \over \delta O}\ri)^2+{1\/2} m^2\conf^{-2\D}O ^2+{1\/2}\conf^{2} \g^{ij}\p_i(\conf^{-\D} O) \p_j( \conf^{-\D} O)+ \cH_{\text{int}} ,
\end{split}
\ee
Here, we have substituted the form \eqref{solansatz} into the Hamiltonian constraint \eqref{hrewritten}. We have explicitly displayed bilinear combinations of terms where a single functional derivative acts on $\solfunc$. Indices are still raised and lowered using $g_{i j}$ and $g^{i j}$ and therefore $ g^{i j}=\conf^{2} \gamma^{i j}$.  We have used $\doubleder$ to indicate the action of second-order functional derivatives on $\solfunc$ and the form of $\doubleder$ can be read off from \eqref{hrewritten}.

The action of the second-order functional derivatives is subtle. This is because the action of a double functional derivative on a local term in $\solfunc$  can generate a divergent $\delta(0)$ term. The precise form of these terms depends on the normal ordering prescription used to define the Hamiltonian constraint and it is reasonable to believe that the $\delta(0)$ terms can be removed by a judicious choice of normal ordering. (See  Appendix \ref{app:contact} for more discussion.)

Fortuitously these terms do not enter the leading-order analysis. This is because the terms in $\doubleder$ are linear in $\solfunc$ whereas the first-order functional derivative terms displayed in \eqref{Hpsi} are quadratic. Since all the terms in \eqref{solansatz} (except for $\gser_0$) grow with $\conf$ the terms in $\doubleder$ always contribute with a lower power of $\conf$. We will see below that the leading contribution from $\doubleder$ can, at most, change the form of the anomaly polynomial at subleading order in $G_N$ but does not change any of the structural features of the answer.

The expression above involves the Ricci scalar of the metric $g$. In terms of the variables, $\conf$ and $\gamma$, this can be written as 
\be
\label{conftransr}
R[g] = {1 \over \conf^2} \left[ R[\g] -2 (d-1) \conf^2 \nabla^i\nabla_i \log(\conf)  + (d - 1) (d - 2)\conf^2 \nabla^i \log(\conf) \nabla_i \log(\conf) \right]~.
\ee
Note that since indices are raised by $g^{ij}$, the terms inside the bracket involving derivatives of $\log (\conf)$ are $\Or[1]$. Therefore we expect that $R$ is of order ${\conf^{-2}}$. In fact, the magnitude of $R$ in cosmological units can be used as an estimate of $\conf^{-2}$ that does not rely on a specific choice of coordinate system.

\subsubsection{Gravitational part}

We now solve the WDW equation order by order in the large-$\conf$ expansion, focusing first on the gravitational part.

\pg{Leading term.} 

The largest term that appears without a derivative in \eqref{Hpsi} is the cosmological constant term. This immediately leads to the conclusion that
\[
\alpha = d
\]
since any larger term in the expansion cannot be cancelled in \eqref{Hpsi}. The leading term in the WDW equation gives
\be
\label{leadingwdw}
{2\k^2\/\Om^{2d}} {1\/4d(d-1)} \le( \Om{\d \gser_d\/\d\Om}\ri)^2 = {\L\/\k^2} ~.
\ee
Using \eqref{ccvalue}, this leads to the equation
\be\label{Xdequation}
\Om {\d \gser_d\/\d\Om} = \pm  {d(d-1)\/\k^2}\Om^d~.
\ee
This yields
\be\label{adsolution}
\gser_d =  -{d-1\/\k^2}\int d^d x\sqrt{g} ~,
\ee
where we have chosen the negative sign for physical reasons explained below, and rewritten the expression in terms of the original variables to make manifest its diffeomorphism invariant form.

We would like to make a few comments.
\begin{enumerate}
\item
We neglected a possible contribution from terms that involve ${\delta \over \delta \gamma_{i j}}$  in going from \eqref{Hpsi} to \eqref{leadingwdw}. It may be checked that no diffeomorphism invariant $\gamma$-dependent term can be added to \eqref{adsolution} while keeping the right hand side of \eqref{leadingwdw}, which is independent of $\gamma$, unchanged.
\item
The choice of negative sign in \eqref{adsolution} corresponds to the fact that we wish to study an expanding de Sitter universe at late times \cite{Susskind:2007pv}. With this sign, the leading part of the wavefunctional satisfies 
\be
\pi^{i j} \Psi \xrightarrow[\conf\to\infty]{} {-i}  {\delta \over \delta g_{ij}}e^{i \gser_d} = -{d - 1 \over 2 \kappa^2}\sqrt{g} g^{i j}  \, e^{i \gser_d}~.
\ee
This is precisely the relation between the canonical momentum and the metric at late times in an expanding universe. A choice of positive sign in \eqref{adsolution} is allowed but would correspond to the part of the wavefunctional that describes a contracting universe.
\item
As advertised, $\gser_d$ is local. This property arises because the right hand side of \eqref{Xdequation} is a number, which does not allow any nonlocal contributions. 
\item
$\gser_d$ is real, which corresponds to an oscillatory phase in the wavefunctional. So although $\gser_d$ has the highest scaling with $\conf$, it does not contribute when the wavefunctional is squared to study expectation values \cite{dssecond2023}.
\end{enumerate}

At the next order, matching orders in $\conf$ we now find that
\be
\label{gserd1eq}
{\delta \gser_{d-1} \over \delta \conf} = 0 \implies \gser_{d-1} = 0.
\ee

Beyond the universal terms, $\gser_d$ and $\gser_{d-1}$, the form of the solution varies slightly in different dimensions. We explain the different cases for low dimensions and the general pattern below. In each case, we will use the following observation. We ignore any term that decays when  $\Om\to+\infty$ in \eqref{solansatz}. When combined with the contribution from $\gser_{d}$, such a term can yield a contribution that decays like ${\conf^{d-1}}$ in \eqref{Hpsi}. Therefore any term in \eqref{solansatz} that yields a contribution of the same order when inserted in \eqref{Hpsi} will be undetermined in our procedure since we expect that its contribution to \eqref{Hpsi} can be cancelled by an appropriate choice of the remainder term.

\paragraph{$\mathbf{d = 2}$.}
In $d=2$, we find that $\gser_{d-2}=\gser_0$ must obey the equation
\be
\label{a0confeqn}
\conf {\delta \gser_0 \over \delta \conf} = {1\over 2 \kappa^2} \conf^2 R \qquad (d=2)~.
\ee
When rewritten in terms of the original variables, this takes on the familiar form
\be
\label{a0eqn}
\left({2}g_{ij} {\delta \over \delta g_{ij}} - \Delta \chi {\delta \over \delta \chi} \right) e^{i \gser_0}  = \cA_2 \,e^{i \gser_0} \qquad (d=2)~ ~,
\ee
with
\be
\label{anomalyd2}
\cA_2= {i \over  2\kappa^2}\sqrt{g}  R ~.
\ee
We note that \eqref{a0eqn} is analogous to the trace anomaly equation for the partition function of a 2d CFT except that the central charge is {\em imaginary} with $c ={12 \pi i \over \kappa^2}$. This can be thought of as the Brown-Henneaux \cite{Brown:1986nw} central charge in AdS but analytically continued to dS as proposed in \cite{Maldacena:2002vr}.

We note some interesting features of \eqref{a0eqn} that will carry over to other dimensions.
\begin{enumerate}
\item
Note that \eqref{a0eqn} does not have a unique solution.  The addition of any term that is independent of $\conf$ to an existing solution of  \eqref{a0confeqn} yields another solution. (In terms of partition functions, this is simply the observation that the trace anomaly equation does not uniquely fix the CFT partition function but only the central charge.)  The reader might wonder why this freedom appears for $\gser_0$ but not for $\gser_{d-1}$. In fact, it is possible to add a nontrivial functional independent of $\conf$ to the solution in \eqref{gserd1eq} and obtain another solution, but such a functional is ruled out by the asymptotic scaling requirement \eqref{asympscaling}.
\item
The form of $\cH_{\text{int}}$ does not place any constraints on $\gser_0$ through \eqref{Hpsi}.  This is for the following reason. The $\Or[\conf^{-1}]$ remainder in $\solfunc$ yields a contribution to \eqref{Hpsi} that is $\Or[\conf^{-d-1}]$ (Such a contribution arises when one ${\delta \over \delta \conf}$ derivative acts on $\gser_d$ and the other acts on the remainder.) Since we are not keeping track of this remainder, we can consistently neglect terms that decay  $\Or[\conf^{-(d+1)}]$ in \eqref{Hpsi}. But in $d=2$, higher-derivative terms decay at least as fast as ${1 \over \conf^3}$. 
\item
The anomaly may receive a possible correction through the action of double derivative terms on $\gser_d$. After accounting for the leading ${1 \over \conf^{2 d}}$ factor in \eqref{Hpsi}, such terms contribute at the same order as $\gser_0$.  The correction is  $\Or[\kappa^0]$ and so it is subleading compared to \eqref{anomalyd2} and its precise value depends on the choice of normal ordering. In Appendix \ref{app:contact}, we show how this term can be made to vanish by a specific choice of normal ordering.
\end{enumerate}

\paragraph{$\mathbf{d = 3}$.}

For $d  \geq 3$,  $\gser_{d-2}$  is determined from the subleading term in the WDW equation, which takes the form
\be
{-}{2\k^2\/\Om^{2d}} {2\/4d(d-1)} \le( \Om{\d \gser_d\/\d\Om}\ri)\le( \Om{\d \gser_{d-2}\/\d\Om}\ri) -{1\/2\k^2}R =0 ~.
\ee
This gives the equation
\be
 \Om{\d \gser_{d-2}\/\d\Om} ={1\/2\k^2} \sqrt{g} R~.
\ee
For $d\geq 3$, this can be integrated to give
\be
\label{gserdm2sol}
\gser_{d-2} ={1\/2(d-2)\k^2} \int d^d x\sqrt{g} R~ ~,
\ee
which can be checked to be a solution using that $\Om {\d\/\d\Om}= 2 g_{ij}{\d\/\d g_{ij}}$.

The next term  $\gser_0$ is unconstrained and can be any function of $\gamma$ that is independent of $\conf$. The action of the differential operators in \eqref{Hpsi} on such a function does yield non-zero terms. However such terms can be cancelled by an appropriate choice of the remainder term in $\solfunc$.  In terms of the original variables, this means that $\gser_0$ is a {\em Weyl invariant function} and we have
\be
\conf {\delta \over \delta \conf}  \gser_0 =0 \qq (d=3)~ ~,
\ee
or, in terms of the original variables,
\be
\left(2 g_{ij}{\d\/\d g_{ij}} - \Delta {\delta \over \delta \chi} \right)e^{i \gser_0} =0 \qq (d=3)~.
\ee
Once again, it is not necessary to keep track of interactions since they do not contribute to the non-decaying parts of $\solfunc$ by the same power-counting argument that was given for $d=2$.

\paragraph{$\mathbf{d \geq 3}$.} The procedure outlined above can be continued to any dimension. The pattern is that higher order terms are determined by the recursive equation coming from
\be\label{eqHigher}
 {2\k^2\/\Om^{2d}}\le[  {1\/4d(d-1)}\le(\Om{\d \cF \/\d\Om} \ri)^2-g_{ik}g_{j\l}(\dgtr^{ij}\cF)( \dgtr^{k\l}\cF) \ri] = \cH_\r{int}~ ~,
\ee
where the power-counting argument above tells us that $\cH_\r{int}$ can contribute to higher-order terms.  Here we have rewritten the metric variations in terms of the traceless metric variation
\be
\dgtr^{ij} = {\d\/\d g_{ij}}-{1\/d} g^{ij}g_{kl}{\d\/\d g_{k\l}}~ ~,
\ee
using the identity \eqref{deltagammaidentity}. The term of order $d-2n$ is determined from this equation as 
\be
\begin{split}
&\Om {\d\/\d\Om} \gser_{d-2n} \\ &= -{2\k^2\/\Om^d} \sum_{k=1}^{n-1}\le( g_{ik}g_{j\l}(\d_g^{ij}\gser_{d-2 k})( \d_g^{k\l}\gser_{d-2 (n-k)}) - {1\/4d(d-1)}\le( \Om {\d \gser_{d-2k}\/\d\Om} \ri)\le(\Om {\d \gser_{d-2(n-k)}\/\d\Om}\ri)\ri) \\
&+ \Om^d \cH_\r{int}~.
\end{split}
\ee
This recursive structure determines all the higher order terms from the first two $\gser_d$ and $\gser_{d-2}$.\footnote{Note that a similar recursive structure was also observed in \cite{Papadimitriou:2004ap} in the context of holographic renormalization. }

As we have pointed out in $d=2$ and $d=3$, since the equation involves functional derivatives with respect to $\conf$, it never fixes $X_0$ uniquely. Given one solution, there is always the freedom to add an $\conf$-independent functional to $X_0$. 

We illustrate the procedure with the next term $\gser_{d-4}$ in pure Einstein gravity where $\cH_\r{int}=0$. The above equation gives
\be
\Om{\d \gser_{d-4} \/\d\Om}=-{1\/2(d-2)^2\k^2}\sqrt{g}\le( R_{ij}R^{ij}-{d\/4(d-1)}R^2\ri)~ ~,
\ee
as derived in Appendix \ref{app:higherorder}.

In $d=4$, we obtain the equation
\be
\Om {\d \gser_0\/\d\Om} = -{1\/8\k^2} \sqrt{g}\le( R_{ij}R^{ij}-{1\/3}R^2\ri) ~,
\ee
which leads to 
\be
\left(2g_{ij}{\d\/\d g_{ij}}  - \Delta \chi {\delta \over \delta \chi} \right) e^{i \gser_0} = \cA_4\, e^{i \gser_0} ~,
\ee
where we have defined
\be
\label{anomalyd4}
\cA_4 \equiv  -{i\/8\k^2} \sqrt{g}\le( R_{ij}R^{ij}-{1\/3}R^2\ri)~.
\ee
We recognize the trace anomaly equation for a four-dimensional CFT partition function $Z = e^{i\gser_0}$. The anomaly can be written as
\be
\cA_4  ={1\/16\pi^2}\sqrt{g}(-a E_4+c W_{abcd}W^{abcd}) ~,
\ee
using the Euler density and Weyl squared curvature given in \eqref{EulerWeyl} with the coefficients
\be
a=c= -{i \pi^2\/ \k^2} = - {i\pi\/8G_N}~.
\ee
This is, up to the factor of $-i$, the anomaly of a holographic CFT$_4$ obtained using holographic renormalization in AdS$_5$ \cite{Balasubramanian:1999re}. It may be checked that this anomaly polynomial receives corrections at subleading order in $\kappa$  from higher-derivative terms and also from a choice of normal ordering in the WDW equation. 

In $d\geq 5$, this equation can be integrated to give
\be
\label{gserdm4sol}
\gser_{d-4} = -{1\/2(d-2)^2(d-4)\k^2}\int d^dx\,\sqrt{g}\le( R_{ij}R^{ij}-{d\/4(d-1)}R^2\ri)~.
\ee
For $d=5$ , the formula above completes the gravitational part of the series.  $X_0$ can be any functional independent of $\conf$. For higher $d$, one must continue the expansion above until $X_0$.

\subsubsection{Matter part}

We now solve for the matter part. 

\pg{Leading term.} The leading term $\matser_\b$ is determined from the equation
\be
{-}{2\k^2\/\Om^{2d}} {2\/4d(d-1)}  \le( \Om{\d \gser_d\/\d\Om}\ri)\le(\Om{\d \matser_\b\/\d\Om}+\D  O {\d \matser_\b\/\d O}\ri)+{1\/2}\Om^{2(\D-d)}\le({\d \matser_\b\/\d O} \ri)^2  + {1\/2}m^2 \Om^{-2\D}O^2 =0~. 
\ee
A solution is only possible if these terms compete which requires
\be
\b=d-2\D~ ~,
\ee
and leads to the equation
\be
 {2\/\sqrt{g}} g_{ij}{\d \matser_\b\/\d g_{ij}}+{1\/2g }\le({\d \matser_\b\/\d \chi} \ri)^2  + {1\/2}m^2 \chi^2 =0  ~,
\ee
which we have written in terms of the original variables.  The solution takes the form
\be
\matser_\b = b_\b \int d^dx\,\sqrt{g}\chi^2 ~,
\ee
where $b_\b$ satisfies
\be\label{eq:b_equation}
4b_\b^2 +2 db_\b+ m^2=0~.
\ee

\paragraph{Mass formula.} 
We can determine the relation between $b_{\beta}$ and $\Delta$ by examining the classical limit.  In the classical theory, in an asymptotically de Sitter spacetime we have
\be
\pi = g_{ij}\frac{\delta \cS}{\delta{\dot{g}_{ij}}} = -{d (d-1)  \over 2N\kappa^2} \conf^{d-1} \dot{\conf}, \qquad \pi_{\chi} = {\d \cS\/\d\dot\chi} ={1\over N}\conf^{d} \dot{\chi}~ ~,
\ee
where $N$ is the lapse function, and $\cS$ is the Einstein-Hilbert action with possible interactions. 
From the definition of $\Delta$ we expect that
\begin{equation}
	\frac{\dot{\chi}}{\chi} = -\Delta \frac{\dot{\conf}}{\conf}~,
\end{equation}
up to terms that vanish at large $\conf$. Here we have not used the equations of motion but simply the kinematic definition of $\Delta$   in \eqref{rescalechi} which determines the scaling behaviour of the field at large volume. In terms of the corresponding canonical momenta, this equation can be written as
\be
\label{classicalmomrel}
{1 \over \chi} \pi_{\chi} =   {2\kappa^2 \Delta\over d (d-1)} \pi~, \qquad \text{(classical~expectation)}~.
\ee 
We can compare this classical expectation with the relation obtained from our wavefunctional. Using the leading order solution for the matter and metric sector, and the relations $\pi = -i g_{ij} {\d\/\d  g_{ij}} $ and $\pi_\chi = -i{\d\/\d\chi}$ on the wavefunctional, we find
\be
\label{mattlargeomrel}
\begin{split}
&\frac{1}{\chi}\pi_{\chi} \Psi = 2 b_{\beta}\Om^d \,\Psi~, \\
&\pi \Psi =  -{d (d -1) \over 2\kappa^2} \conf^{d} \,\Psi ~,
\end{split}
\ee
up to subleading terms in the $\Om\to+\infty$ limit.   By matching this with \eqref{classicalmomrel}, we see that we must have
\be\label{valbbeta}
b_{\beta} =- {\Delta \over 2}~.
\ee
Substituting this in \eqref{eq:b_equation} we find that $\Delta$ must be related to the mass through
\be\label{massFormula}
\Delta (d - \Delta) = m^2~.
\ee
With these substitutions, the leading term in the matter series becomes
\be
\label{matserbsol}
\matser_{\beta} = -{\Delta \over 2} \int d^dx\,\sqrt{g}\chi^2~.
\ee
Our derivation involved a correspondence with the classical limit, and the choice of specific orderings, such as the ordering of ${1 \over \chi} \pi_{\chi}$ in \eqref{mattlargeomrel}. Therefore the result \eqref{massFormula} can be thought of  as being valid to leading order in $\kappa$.

There is a class of solutions to \eqref{massFormula} corresponding to $m>{d\/2}$ of the form
\be
\D = {d\/2}\pm i\nu,\qq \nu = \sqrt{m^2- {d^2\/4}}~.
\ee
If one studies a nongravitational quantum field theory with this mass, then the single-particle states in such a theory lie in the principal series of representations of the conformal group  $\confgrp$ that is the isometry group of dS$_d$ \cite{Bros:1995js}.  The complementary series correspond to masses in the range $0< m < {d\/2}$ and we can restrict to the range $0< \D< {d/2}$. For a nice recent review, we refer the reader to \cite{Sun:2021thf}.

\pg{Subleading term.} In the principal series, the leading matter term is the only term we have since $\r{Re}\,\b = 0$ so that the subleading term decays as $\Om\to + \infty$. 

In the complementary series, the subleading term $\matser_{\b-2}$ contributes when 
\be
\D \leq {d-2\/2}~.
\ee
It is determined by the equation
\begin{align}
	\begin{split}
	&-\frac{2\kappa^2}{\conf^{2d}} \frac{1}{4d(d-1)} \encsq{2\enc{\conf\frac{\delta \gser_d}{\delta\conf}}\enc{\conf\frac{\delta \matser_{\beta-2}}{\delta\conf} + \Delta O\frac{\delta \matser_{\beta-2}}{\delta O}} + 2\enc{\conf\frac{\delta \gser_{d-2}}{\delta\conf}}\enc{\conf\frac{\delta \matser_{\beta}}{\delta\conf} + \Delta O\frac{\delta \matser_{\beta}}{\delta O}} } \\
	& + \frac{\conf^{2\Delta}}{2 \conf^{2d}}\, 2\enc{\frac{\delta \matser_\beta}{\delta O}}\enc{\frac{\delta \matser_{\beta-2}}{\delta O}} +{1\/2}\conf^{2} \g^{ij}\p_i(\conf^{-\D} O) \p_j( \conf^{-\D} O)= 0.
	\end{split}
\end{align}
Restoring the variables $\chi$ and $g_{ij}$ and substituting in the known higher order functionals gives the simpler looking form
\begin{equation}
	\frac{2}{\sqrt{g}}\enc{g_{ij}\frac{\delta}{\delta g_{ij}} + b_\beta \chi\frac{\delta}{\delta \chi}} \matser_{\beta-2} - \frac{b_\beta }{2(d-1)}R \chi^2 +\frac{1}{2}g^{ij}\partial_i\chi \partial_j \chi=0.
\end{equation}

The solution (derived in Appendix \ref{app:subleadmatter}) is
\be
\label{matserbm2sol}
\matser_{\b-2} =-{1\/2(d-2-2\D)}\int d^dx \sqrt{g}\le(  g^{ij}\p_i \chi\p_j\chi +{ \D\/2(d-1)}R[g] \chi^2 \ri) ~,
\ee
where we have substituted $b_\beta$ from \eqref{valbbeta}.

When $\Delta < {d - 4 \over 2}$, additional terms appear in the matter series and these terms can be worked out recursively using the WDW equation. 

\subsection{Asymptotic solution \label{subsecasympsol}}

We now give the general form of the asymptotic solution in the limit $\conf \rightarrow \infty$. What we showed is that the constraints imply that it takes the form
\be
\label{solform}
\Psi \underset{\conf \to +\infty}{\longrightarrow} e^{i S[g, \chi]}Z[g,\chi]~.
\ee
Here, $S$ is a universal phase factor that comprises integrals of {\em local} densities. It takes the form
\be
\label{sexpansionsummary}
S = \sum_{n=0}^{d-1} \gser_{d-n} + \sum_{m=0}^{\lfloor\r{Re}\,\b \rfloor} \matser_{\beta- m}  ~,
\ee
where $\beta = d - 2 \Delta$  and $\Delta$ is related to the mass of the field by \eqref{massFormula}.  Terms corresponding to odd values of $m$ and $n$ vanish in \eqref{sexpansionsummary}. Explicit  expressions for $n=0,2,4$ are given in \eqref{adsolution}, \eqref{gserdm2sol} and \eqref{gserdm4sol} respectively and for $m = 0, 2$ in \eqref{matserbsol} and \eqref{matserbm2sol} respectively. All terms in $S$ are subject to the momentum constraint and so they are invariant under $d$-dimensional diffeomorphisms.

The factor $Z[g,\chi] = e^{i \gser_0}$ is a diffeomorphism invariant functional involving possibly nonlocal terms in $g$ and $\chi$ and it has simple Weyl transformation properties.\be
\label{zweyltransform}
\conf {\delta \over \delta \conf (x)}Z[g, \chi]  = {\cal A}_{d}[g] Z[g,\chi]~.
\ee
The anomaly ${\cal A}_{d}[g]$ vanishes for $d$ odd. For even $d$, it can be expressed in terms of curvature invariants and explicit expressions for $d = 2$ and $d=4$ are provided in \eqref{anomalyd2} and \eqref{anomalyd4} respectively.  Since $S$ is real and the anomaly is imaginary, the absolute value of the wavefunctional $|\Psi[g,\chi]|^2$ is always  diffeomorphism and Weyl invariant. 

 $Z[g,\chi]$ is not uniquely fixed by the WDW equation. Structurally, this is because the WDW, at large $\conf$, relates functional derivatives of a term with a given scaling in $\conf$ to a source term that arises from terms with higher scaling in $\conf$. Therefore an existing solution for $\gser_{0}$ can be modified by the addition of an $\conf$-independent term (i.e. a Weyl invariant term) to yield another solution. This can also be seen from the fact that the solution to \eqref{zweyltransform} is not unique.

Once $Z[g, \chi]$ has been chosen it is possible to use the equation \eqref{hpconstraint} to continue the series expansion in $\conf$. The choice made for $Z[g,\chi]$ then controls the terms in the wavefunctional that decay with $\conf$. Physically, this can be thought of as follows. We specify a state at late times ($\conf \rightarrow \infty$) by specifying the arbitrary functional in $Z[g,\chi]$.  If we wish to ask questions about finite-time physics, then we must determine the full dependence of the wavefunctional on $\conf$. This dependence is sensitive to the interaction terms that appear in the Hamiltonian constraint, and we do not investigate it in this paper. 

The precise numerical values that we have found for the anomaly polynomials, and for $\gser_{d-n}$ and $\matser_{\beta - m}$ rely on a choice of normal ordering and, in some cases, can be affected by higher-order terms in the interactions. However, the structural properties of the WDW equation at large $\conf$ --- which are that the higher-order terms are fixed by a recursive set of functional equations,  $e^{i X_0}$ is left undetermined up to its Weyl transformation properties, and the momentum constraint imposes diffeomorphism invariance on each term --- are robust. Therefore we conjecture that the {\em form} of the solution \eqref{solform} is valid to all orders in the $\kappa$-expansion.

\pg{Discussion and comparison to AdS/CFT.}
The phase factor $S[g, \chi]$ is closely related to the counterterm-action that arises in holographic renormalization \cite{Henningson:1998gx,Hyun:1998vg,Balasubramanian:1999re,Emparan:1999pm,deHaro:2000vlm,Skenderis:2002wp}. The reason for this can be understood as follows. 

In Euclidean AdS, it is possible to study the action  of the bulk theory, with boundary conditions imposed on radial slices of the spacetime.  The procedure of holographic renormalization identifies the divergences in this on-shell action.  The wavefunctional in dS in the Euclidean vacuum can be obtained by analytically continuing this action  \cite{Maldacena:2002vr}. Since the phase factor is universal, it is sufficient to determine it in a single state, as can be done using this procedure. 

However, the analytic continuation introduces factors of $i$ in the anomaly. Also, it makes the phase factor oscillatory in the dS case,  whereas the counterterm action is real in AdS. The terms that appear in the matter sector are also slightly different in dS. This is because the dimension of operators dual to matter fields is always real in AdS but it can be complex in dS.

Second, the on-shell action on radial slices also obeys the constraints of diffeomorphism-invariance that lead to a WDW-type equation. As one approaches the boundary of AdS, it is possible to solve this equation asymptotically \cite{Freidel:2008sh} (See also \cite{Cianfrani:2013oja}) and this procedure also yields the correct divergent terms.

However, there is an important conceptual difference between the ``radial WDW'' equation and the one that we are studying. The radial wavefunctional is subject to regularity at $r = 0$. This fixes its asymptotic form at $r = \infty$ to be a phase factor times the partition function of a specific CFT --- the CFT that is dual to the bulk AdS theory by the AdS/CFT correspondence \cite{Maldacena:1997re,Witten:1998qj,Gubser:1998bc}. These constraints cannot be seen from the asymptotic analysis in AdS.

From quantum mechanics, we expect that such constraints should {\em not} apply to our solutions. The specification of $Z[g,\chi]$ is akin to the specification of a state on a late-time spatial slice. But, in quantum mechanics, such a specification can be performed freely and is not subject to constraints that come from the time evolution of the state.  No such principle applies in AdS where the ``radial wavefunctional'' specifies data on a timelike boundary, which cannot be done freely.  It is possible that there are additional constraints that restrict the allowed set of $Z[g,\chi]$ \cite{Arkani-Hamed:2017fdk} but they are not evident in our analysis.

\section{Solution space as theory space \label{sechilbastheory}}

In the previous section, we have argued that, in the large-volume limit, all solutions of the WDW equation take the form given in \eqref{solform} --- a universal phase factor multiplied with a diffeomorphism invariant function, $Z[g, \chi]$ with simple Weyl transformation properties. 

Since the phase factor is universal, each distinct choice of $Z[g,\chi]$ leads to a distinct solution of the WDW equation. Second, in quantum mechanics, states can be read off from any time slice --- even if that time slice is at arbitrarily late times. So,  we expect that the large-volume behaviour of the wavefunctional completely specifies its form everywhere. Therefore each distinct state leads to a distinct choice of $Z[g, \chi]$.  The two observations above lead to the conclusion that there is a one-to-one map between the space of allowed $Z[g, \chi]$ and the space of states in the theory.

We now investigate the properties of $Z[g,\chi]$  more carefully and argue that the space of allowed functionals can be thought of as ``theory space''.

\subsection{$Z$ as a CFT partition function }
Now, let us examine the equation \eqref{zweyltransform} together with the momentum constraint as written in terms of the original variables, $g$ and $\chi$.

The relations that we find on $Z$ are the following
\be
\label{Zidentities}
\begin{split}
&\le(2\sqrt{g}\nabla_{i} {1\/\sqrt{g}} {\delta \over \delta g_{i j}} - g^{i j} \p_i\chi{\delta \over \delta \chi} \ri) Z[g, \chi] = 0; \qquad {\text{(diffeomorphism invariance)}} \\
&\le(2 g_{i j} {\delta \over \delta g_{i j}} -  \Delta \chi {\delta \over \delta \chi} \ri) Z[g, \chi]  = {\cal A}_d Z[g, \chi]. \qquad \text{(Weyl transformation)}
\end{split}
\ee
The first equation comes from the momentum constraint and expresses diffeomorphism invariance while the second equation comes from the Hamiltonian constraint and expresses the anomalous Weyl transformation. 

This can be made more explicit by writing the action of an infinitesimal \diffWeyl{} transformation on the metric and the scalar field
 \be
\label{generaldiffweyl}
\d_{(\xi,\vphi)} g_{ij}  =\cL_\xi g_{ij} +2 \vphi g_{ij},\qq \d_{(\xi,\vphi)}\chi = \xi^i\p_i\chi -\D \vphi\chi~.
\ee
Then the equations  \eqref{Zidentities} are equivalent to 
\be\label{Zintanomaly}
\d_{(\xi,\vphi)} \log Z[g,\chi]= \int d^d x\,\vphi(x)\cA_d(x) ~,
\ee
which can be proven by taking  the functional derivatives with respect to $\vphi(x)$ and $\xi_k(x)$.

 The equations \eqref{Zidentities} are also obeyed by the partition function of $d$-dimensional CFT with a source $\chi$ turned on for an operator $\phi$  of dimension $\bar\D = d - \Delta$ on a Euclidean spacetime with metric $g_{i j}$
\be
\label{cftpart}
Z_\r{CFT}[g,\chi] = \langle e^{-\int d^d x\,\chi \phi} \rangle_{g_{ij}}~,
\ee
which obeys \eqref{Zidentities} with an appropriate choice of ${\cal A}_d$.

However, several cautionary remarks are in order.
\begin{enumerate}
\item
First, the anomaly polynomial that appears for even $d$ in \eqref{Zidentities} is imaginary.  Second, the dimension of $\phi$ can be complex for sufficiently large mass.  This can be seen from the mass-dimension relation \eqref{massFormula}. 
\item
Second, it is possible to obtain correlation function of the stress-tensor and of the operator $\phi$ by differentiating $Z_\r{CFT}$. In a local CFT, such correlators obey various constraints, including  the constraints of cluster decomposition that follow from locality. Our analysis does not provide any reason to believe that the quantities obtained by functional differentiation of $Z$ with respect to the metric or $\chi$ should obey such constraints.  
\item
Relatedly, the space of allowed $Z$'s has a natural vector-space structure since this space is the space of states for a quantum-mechanical system.  But a vector-space structure is unnatural in the space of CFT partition functions since the linear combinations of two partition functions of local CFTs does not, in general, correspond to the partition function of any other local CFT.
\end{enumerate}
Therefore, although $Z[g,\chi]$ obeys the same equations that are obeyed by a CFT partition function, it does not necessarily correspond to the partition function of a unitary or local CFT.

\subsection{Coefficient functions as CFT correlators}

We will now expand $\log\,Z[g,\chi]$ in the metric and matter fluctuations.
This will give a basis of functionals for the solution space, comprising those $Z[g,\chi]$ that do not vanish in the limit where $g_{i j} \rightarrow \delta_{i j}$ and $\chi \rightarrow 0$.\footnote{When we introduce a norm on solutions \cite{dssecond2023}, it will turn out that such solutions do not yield normalizable states. For the present analysis, this issue is not relevant.  In the next section, we will study a different basis corresponding to functionals that vanish in the limit  $g_{i j} \rightarrow \delta_{i j}$ and $\chi \rightarrow 0$. Those functionals are linear combinations of the functionals studied here, and provide a normalizable basis for the Hilbert space.}

\subsubsection{Weyl transformation of the variables \label{weyltophysical}}

So far we have considered the wavefunctional in the limit where the volume of the spatial slice becomes arbitrarily large.  Physically, we are interested in studying fluctuations about an asymptotically de Sitter spacetime, where the metric on a spatial slice takes on the form 
\be
\label{gphys}
g_{ij}^\text{phys} = {4 \omega^2  \over (1 + |x|^2)^2} (\d_{ij}+\k h_{ij}) ~,
\ee
which is a perturbation of the round metric on $S^{d}$ in coordinates $x^i$ rescaled with a large Weyl factor $\omega$.

It may be seen from \eqref{Zidentities} that $|\Psi[g,\chi]|^2 = |Z[g, \chi]|^2$ is diffeomorphism and Weyl invariant. The Weyl anomaly is imaginary so we have
\be
 \le[2g_{i j} {\delta \over \delta g_{i j}} -  \Delta \chi {\delta \over \delta \chi} \ri] (Z[g, \chi] Z[g, \chi]^*) = \big({\cal A}_d Z[g, \chi] \big) Z[g, \chi]^*  + \text{c.c} = 0 ~,
\ee
using that ${\cal A}_d^* = -{\cal A}_d$.

Therefore, at the cost of an additional phase in the wavefunctional in even dimensions, we can make a Weyl transformation of the physical fields  and study the behaviour of $Z[g, \chi]$ in the regime where 
\be
\label{gnearflat}
g_{i j} = \delta_{i j} + \kappa h_{i j}~.
\ee
In converting the physical metric \eqref{gphys} to the form above, we have not only removed the large factor $\omega(x)$ but also made use of the fact that the round metric is related by a Weyl transformation to the flat metric. This does not change the fact that the spatial slices are topologically $S^{d}$ and in \cite{dssecond2023}, we will utilize this when we place boundary conditions at $|x| \to \infty$.

The Weyl transformation that takes the physical metric $g_{ij}^\text{phys}$ to $g_{ij}$ also rescales the matter fields according to
\be
\label{matterphysrescaled}
\chi = \le({2 \w \over 1 + |x|^2}\ri) ^{\D} \chi^\text{phys}~.
\ee
We will now study an expansion of $Z[g, \chi]$ in powers of $\chi$ and $h$ as it appears in \eqref{gnearflat}.  This is a convenient regime in which to study $Z$. If the value of $Z$ is required in the physical regime, the phase factor in the wavefunctional can always be worked out using the anomaly equation and undoing the transformation from \eqref{gnearflat} to the original physical metric \eqref{gphys}.

\subsubsection{Expansion of $Z$ \label{subsecexpzg}}

In what follows, it is convenient to introduce some notation. We write
\be\label{Zexpanded}
Z[g,\chi] = \exp\left[\sum_{m,n} \kappa^n \coeff[n,m] [h, \ldots h, \chi, \ldots \chi] \right] ~,
\ee
where we have defined multi-linear functionals, $\coeff[n,m]$  that take $n$ tensor fields and $m$ scalar fields as input and return a $c$-number. 
\be
\label{coeffincoeff}
\begin{split}
&\coeff[n,m][h^{(1)}, \ldots h^{(n)}, \chi^{(1)}, \ldots \chi^{(m)}] \\
&\equiv {1 \over n! m!} \int d \vec{y}d\vec{z} \, G_{n,m}^{\vI\vJ}(\vec{y}, \vec{z}) h^{(1)}_{i_1 j_1}(y_1) \ldots h^{(n)}_{i_n j_n}(y_n) \chi^{(1)}(z_1) \ldots \chi^{(m)}(z_m) ~.
\end{split}
\ee
The ``coefficient functions'' $G_{n,m}^{\vI\vJ}(\vec{y},\vec{z})$ depend on $\vec{y} = (y_1,\dots,y_n), \vec{z}=(z_1,\dots,z_m)$ and are tensors with multi-indices $\vI = (i_1,\dots,i_n), \vJ=(j_1,\dots,j_n)$ that are symmetric in the $(i_a, j_a)$ indices.  Here $\vec{x} = (\vec{y}, \vec{z})$ is a collective symbol for all the coordinates in the equation. We demand that these functionals $\coeff[n,m]$ be symmetric under the interchange of any two of the $h^{(k)}$ or any two of the $\chi^{(k)}$, which means that $G_{n,m}^{\vec{i} \vec{j}}$ are symmetric under interchange of any two $z$ coordinates and the simultaneous interchange of any two $y$-coordinates and the associated tensor indices.

We will now use the relations \eqref{Zidentities} to derive constraints on the functions $\coeff[n,m]$. A similar analysis was performed in \cite{Pimentel:2013gza} for the Euclidean vacuum. (See also \cite{Hartle:2008ng}.) The general strategy that we adopt will be the following. Under an infinitesimal \diffWeyl{} transformation \eqref{generaldiffweyl}, $h_{ij}$ transforms as
\be
\label{htransform}
\delta_{\xi,\vphi}h_{ij} =  H_{ij} + {1\/\k} I_{ij} ~,
\ee
where $H_{ij} = \cL_\xi h_{ij}+2\vphi h_{ij}$ is a piece linear in $h_{ij}$  and  $I_{ij} = \p_i\xi^k \d_{jk} +\p_j\xi^k \d_{ik} + 2\vphi\d_{ij}$ is an inhomogeneous piece that comes from the transformation of the background metric.
 
We then have the variation 
\be
\label{dellogzgeneral}
\begin{split}
\diff \log Z &= \sum_{n,m}  \kappa^{n}(\diff \log Z)_{n,m},
\end{split}
\ee
where we have collected terms according to the expansion in $\k$:
\be
\begin{split}
(\diff \log Z)_{n,m} \equiv  &(n+1)\coeff[n+1,m][I, h, \ldots h, \chi, \ldots \chi] + n  \coeff[n,m][H, h, \ldots h, \chi, \ldots \chi] \\
 &+ m  \coeff[n,m] [h, \ldots h, \delta \chi, \ldots \chi]~.
\end{split}
\ee
The constraint \eqref{Zintanomaly} then leads to identities that relate $\coeff[n+1,m]$ to $\coeff[n,m]$. These identities are derived in Appendix \ref{app:Ward}. 
   As a consequence of the anomalous Weyl transformation, we obtain a ``trace identity'' 
\be\label{traceidentity}
2 \d_{ij}G_{n+1,m}^{ij \vI \vJ}(u,\vec{y},\vec{z}) = \le(-2 \sum_{a=1}^n \d^{(d)}(u-y_a)+\D \sum_{b=1}^m \d^{(d)}(u-z_b) \ri)G_{n,m}^{\vI \vJ}(\vec{y},\vec{z}) +  \d_{m,0}\cA_d^{\vI \vJ}(u,\vec{y})~ ~,
\ee
where  
\be
{\cal A}_{d}^{\vI \vJ}(u,\vec{y}) = {1 \over \kappa^n} {\delta^n \over \delta h_{i_1 j_1}(y_1) \ldots  \delta h_{i_n j_n}(y_n)}\cA_d(u) ~,
\ee
is an ultra-local term obtained from the expansion of the anomaly $\cA_d$ in the fluctuation, which only appears for even $d$.

The invariance under diffeomorphisms leads to a ``divergence identity''.
\bea\label{dividentity}
&& 2  \d_{jk}\p_i G_{n+1,m}^{ij\vI \vJ}(u,\vec{y},\vec{z})=-\sum_{b=1}^m {\p\/\p z_b^k} \Big[ \d^{(d)}(u-z_b)G_{n,m}^{\vI \vJ}(\vec{y},\vec{z})\Big]\-
&& +\sum_{a=1}^n \Big[ -{\p\/\p y_a^k} \big[ \d^{(d)}(u-y_a) G_{n,m}^{\vI \vJ}(\vec{y},\vec{z}) \big]+ G_{n,m}^{\vI' \vJ'}(\vec{y},\vec{z})\big( \d_{i_a'}^{i_a}\d^{j_a}_k {\p\/\p y^{j_a'}}+ \d_{j_a'}^{j_a} \d^{i_a}_k {\p\/\p y^{i_a'}}\big)\d^{(d)}(u-y_a)\Big]~,
\eea 
where in the bracketed expression, we use $\vI'$ and $\vJ'$ to denote the multi-index where $(i_a,j_a)$ has been replaced by $(i_a',j_a')$ for the current $a$ in the sum. (See \eqref{ipjpdef}.)

\subsubsection{Conformal symmetry of the coefficient functions}

A special role is played by the combinations of diffeomorphism and Weyl transformations that leave the background flat metric invariant. These are conformal transformations. Under these transformations, the inhomogeneous piece in \eqref{htransform} vanishes: 
\be
I_{ij} = \p_i \xi^k \d_{jk}+ \p_j\xi^k \d_{ik} - {2\/d}\d_{ij}\p_k\xi^k =0~ ~,
\ee
and corresponds to taking $\xi$ to be a conformal Killing vector.

This imposes conformal invariance on the functions $G$. More specifically under a conformal transformation $\vec{y} \rightarrow \vec{y}',\vec{z}\ra \vec{z}'$, we have
\be\label{finiteConf}
G^{\vI \vJ}_{m,n}(\vec{y}',\vec{z}')= \Big(\prod_{a=1}^n R_{i_a'}^{i_a}(y_a) R^{j_a}_{j_a'}(y_a) \Lambda(y_a)^d\Big)\Big(\prod_{b=1}^m \Lambda(z_b)^{d-\D}\Big) G_{m,n}^{\vI '\vJ'}(\vec{y},\vec{z}) ~,
\ee
where 
\be
R^{i}_{i'}(x)= \Lambda(x) J^{i}_{i'}(x),\qq J^{i}_{i'}(x)={\p x^{i}\/\p x^{i'}} ,\qq \Lambda(x) = |\r{det} \,J(x)|^{-1/d}~ ~,
\ee
see Appendix \ref{app:Wardconf} for the derivation.

This shows that the coefficient functions $G_{n,m}^{\vI \vJ}$ obey the same identities as connected CFT correlators. We can write
\be
\label{gnmascorr}
G_{n,m}^{\vI \vJ}(\vec{y}, \vec{z}) \sim \langle T^{i_1 j_1}(y_1) \ldots T^{i_n j_n}(y_n) \phi(z_1) \ldots \phi(z_m) \rangle^{\text{connected}}_\r{CFT} ~,
\ee
where $T^{i j}$ is an operator of spin $2$ and dimension $d$ and $\phi$ is an operator of dimension $d - \Delta$.\footnote{From \eqref{Zexpanded} and \eqref{coeffincoeff} it may be seen that the ``correlators'' differ from the conventional correlators, $\langle \ldots T^{i j} \ldots \phi \ldots \rangle_{\text{conv}} = [\ldots {1 \over \sqrt{g}} {\delta  \over \delta g_{i j}}\ldots {1 \over \sqrt{g}} {\delta \over \delta \chi} \ldots ]\log(Z)$,  since they are defined without a factor of ${1 \over \sqrt{g}}$.  So the contact terms that appear in our Ward identities are slightly different even when $g_{i j} = \delta_{i j}$.}  The reason we put the subscript ``connected'' is because $G_{n,m}$ are obtained by functional differentiation of the logarithm of $Z$. The reason we write $\sim$ rather than equality is to indicate that the similarity between the two sides of \eqref{gnmascorr} is only restricted to the fact that both sides obey the same Ward identities. We reiterate that \eqref{gnmascorr} should be interpreted cautiously beyond this shared property.

\subsection{Solution space and theory space}

We have therefore reached the following conclusion. Say that we are given a set of functions,
\be
\label{listcorr}
\{G_{n,m}^{\vI \vJ}(\vec{y}, \vec{z})\}~,
\ee
for all values of $n$ and $m$, which satisfy the Ward identities \eqref{traceidentity} and \eqref{dividentity} and transform under conformal transformations as \eqref{finiteConf}. Such a list of functions uniquely specifies a valid solution to the WDW equation when assembled together through \eqref{Zexpanded}.

Such a list can also be thought of as defining a ``theory'' with the caveats mentioned above: this theory is a CFT but need not be unitary or local. Moreover, the list of correlators \eqref{listcorr} does not make reference to other operators in the theory beyond those that correspond to fields in the physical spacetime. In this generalized sense, the space of solutions to the WDW equation is like  ``theory space''.

\subsubsection{Relation to the set of  Hartle-Hawking wavefunctionals}

The Hartle-Hawking no boundary proposal \cite{Hartle:1983ai} provides a recipe of computing the wavefunctionals that constitute the solution space. Hartle and Hawking proposed that the vacuum wavefunctional should be computed by performing the Euclidean path integral on a manifold with a single boundary. An alternative technique is to compute the path-integral with boundary sources turned on for the same bulk theory in AdS and then continue the answer to dS \cite{Maldacena:2002vr}. The latter technique allows for the easy inclusion of perturbative corrections to the wavefunctional through the computation of AdS correlators.\footnote{As emphasized in \cite{Harlow:2011ke}, AdS correlators continue to the coefficient functions $G_{n,m}^{\vI \vJ}$ in \eqref{coeffincoeff} and not to correlators on the late-time boundary of dS. These latter correlators are called cosmological correlators and are discussed further in \cite{dssecond2023}. They must be computed by further squaring and integrating the wavefunctional (see \cite{Ghosh:2014kba} for an example) or by means of the in-in formalism \cite{Weinberg:2005vy}.}

This computation produces a wavefunctional that we can denote by $\Psi_{0}[g, \chi]$ and which satisfies the Wheeler-DeWitt equation. It explicitly has the general form we have deduced above; the phase $S$ is the analytic continuation of the divergent part of the on-shell action in AdS and the wavefunctional is obtained by multiplying the phase factor with $Z_0[g,\chi]$ --- the analytic continuation of the partition function of the boundary CFT. Since the details of the coefficient functions that enter the partition function depend on the  bulk Lagrangian, $L_{\text{bulk}}$, we can represent this entire process schematically as
\be
\label{lbulktopsizero}
L_{\text{bulk}} \longrightarrow \Psi_{0}[g, \chi]~.
\ee

The prescription \eqref{lbulktopsizero} leads to an interesting observation. Consider a different bulk Lagrangian, $\widetilde{L}_{\text{bulk}}$ but one which gives rise to the same phase factor and therefore has the same holographic anomaly. It is possible to compute a second wavefunctional using this Lagrangian:
\be
\label{lbulktopsizerotilde}
\widetilde{L}_{\text{bulk}} \longrightarrow \widetilde{\Psi}_{0}[g, \chi]~.
\ee

But since the coefficient functions inside $\Psi_{0}[g, \chi]$ and $\widetilde{\Psi}_{0}[g,\chi]$ satisfy the same Ward identities, both wavefunctionals are valid states in either bulk theory.  The Hartle-Hawking wavefunctional computed for the Lagrangian $\widetilde{L}_{\text{bulk}}$ can be thought of as an ``excited state'' in the theory where the bulk interactions are specified by $L_{\text{bulk}}$. Conversely, if one thinks of the bulk theory with the Lagrangian $\widetilde{L}_{\text{bulk}}$ then it is $\widetilde{\Psi}_{0}[g, \chi]$ that is the vacuum wavefunctional and $\Psi_{0}[g,\chi]$ that is an excited state.

Therefore the space of states contains the set of Hartle-Hawking wavefunctionals for all possible bulk interactions that give rise to the same holographic anomaly.  If one considers a specific bulk theory, then this picks out a specific vector in this space as the one corresponding to the vacuum. But the wavefunctionals for other bulk interactions are still in the state space; they just correspond to non-vacuum states.

\section{The ``small fluctuations'' basis for the Hilbert space \label{secpert}}
In this section, we describe an alternate basis for solutions to the WDW equation at late times that is particularly convenient in the limit where $G_{N} \to 0$.  Although we use the adjective ``small fluctuations'', this basis spans the entire Hilbert space. We show, using this basis, that as $G_N \to 0$, the space of states we have constructed coincides precisely with the Hilbert space constructed by Higuchi \cite{Higuchi:1991tm, Marolf:2008it,Marolf:2008hg}. However, our construction also provides a procedure to systematically correct Higuchi's construction at nonzero $G_N$. 

The basis we introduce in this section has the additional advantage that it will yield normalizable states in the Hilbert space \cite{dssecond2023}.

\subsection{Basis of  ``small fluctuations''}
In section \ref{secsol} and \ref{sechilbastheory} it has been shown that a general solution to the WDW equation is spanned by wavefunctionals of the form 
\be
\label{basestate}
\Psi[g,\chi] = e^{i S} \r{exp}\Big[\sum_{n,m} \kappa^n \coeff[n,m]\Big] ~,
\ee
where, as above,  each $\coeff[n,m]$ takes the form of a conformally invariant ``coefficient function'' integrated with the fluctuations of the metric and the matter fields. The coefficient functions must obey a set of Ward identities and can be identified as correlation functions of $n$-insertions of the ``stress tensor'' and $m$-insertions of a scalar operator with dimension $d-\Delta$ in a non-unitary conformal-field theory.

Here, and in what follows, we do not display the arguments of $\coeff[n,m]$ to condense the notation. It is understood that all $n$ tensor arguments correspond to the metric fluctuation $h_{i j}$ and all $m$ scalar arguments correspond to the matter fluctuation $\chi$.

Consider a state with a specific choice of functionals, $\coeff[n,m]$.  Now consider another set of functionals $\tcoeff[n,m]$, which also satisfy the Ward identities of section \ref{secsol}. Then the combination 
\be
\label{gdeform}
\lcoeff[n,m] = (1-\lambda) \coeff[n,m] + \lambda \tcoeff[n,m] ~,
\ee
also satisfies the identities of section \ref{secsol}. The linear combination chosen above ensures that $\lcoeff[n,m]$ satisfies the Ward identities with the same trace anomaly term. Therefore, the wavefunctional
\be
\Psi_{\lambda}[g, \chi] = e^{i S} \r{exp}\Big[ \sum_{n,m} \kappa^n \lcoeff[n,m]\Big] ~,
\ee
also satisfies the WDW equation asymptotically. Since the solution space is linear this means that 
\be
\label{psiderstate}
\left. {\partial \Psi_{\lambda}[g, \chi] \over \partial \lambda}\right|_{\lambda = 0} =\sum_{n,m} \kappa^n \big(\tcoeff[n,m] - \coeff[n,m] \big) \Psi[g,\chi] ~,
\ee
is also a valid state. The combination above will appear frequently and so we define the notation
\bea
\dcoeff[n,m] &\equiv&  \tcoeff[n,m] -\coeff[n,m] \-
\= {1 \over n! m!} \int   d \vec{y} d \vec{z} \, \delta G^{\vI \vJ}_{n,m}(\vec{y}, \vec{z}) h_{i_1j_1}(y_1) \ldots h_{i_nj_n}(y_n) \chi(z_1) \ldots \chi(z_m) ~.
\eea

We can think of the states \eqref{psiderstate} as corresponding to ``small fluctuations'' about the base state $\Psi[g,\chi]$. Nevertheless, states of the form \eqref{psiderstate} provide a complete basis for the Hilbert space provided we consider all possible changes $\dcoeff[n,m]$. We can refer to this as the ``small fluctuations'' basis for the Hilbert space.

The construction above can be performed about any base state but to make contact with the existing literature we will, henceforth, choose the base state in \eqref{basestate} to be the Hartle-Hawking state, $\Psi_{0}[g,\chi]$. The basis above then naturally corresponds to the basis of fluctuations about the Euclidean vacuum.

A few comments are in order.
\begin{enumerate}
\item
Naively, it might appear possible to take the functionals $\coeff[n,m]$ and $\tcoeff[n,m]$ to coincide for all value of $n,m$ except for some particular values of $n=n_0, m= m_0$. However, this is not possible as both sets of functionals must satisfy the Ward identities. This relates the longitudinal components and the trace of  $\tcoeff[n+1,m]$ to $\tcoeff[n,m]$ for each $n,m$ by equations \eqref{traceidentity} and \eqref{dividentity}. Therefore
\be
\label{lowermeanshigher}
 \dcoeff[n,m] \neq 0 \quad\Rightarrow\quad   \dcoeff[n+1,m] \neq 0~.
\ee
\item
Nevertheless, note that the right hand side of \eqref{lowermeanshigher}, $\dcoeff[n+1,m]$ is not completely fixed by the left hand side, $\dcoeff[n,m]$ This is because the Ward identities only fix the longitudinal components and the trace in $\tcoeff[n+1,m]$. Except for $n+m \leq 2$ there are an infinite number of possible ways to satisfy the Ward identities.
\item
On the other hand, the Ward identities do not prevent the possibility that $\dcoeff[n,m]=0$ for $n < n_0,m < m_0$ for some choice of $n_0, m_0$ but that $\dcoeff[n,m] \neq 0$ for other values of $n,m$.
This might appear slightly puzzling since, in a CFT, all higher-point functions are fixed by three-point functions. However, the coefficient functions can be thought of as  correlators of ``stress-tensor'' and the operators dual to the matter fields.   The wavefunctional does not directly contain terms that correspond to correlators of other primary operators. Within this restricted class of correlators, it is usually possible to change  higher-point correlation functions without changing lower-point functions although some correlators such as the three-point function of a stress-tensor and two scalars are completely fixed in terms of lower-point functions \cite{Osborn:1993cr}.
\end{enumerate}

\subsection{The nongravitational limit \label{nongravstates}} 
The basis above is particularly convenient in the nongravitational limit. When $\kappa \rightarrow 0$, the constraints imposed by the Ward identities become trivial. As we show below, this allows us to obtain a  precise correspondence with Higuchi's basis \cite{Higuchi:1991tm} of dS-invariant states.

Consider two sets of functionals $\dcoeff[n,m]$ and $\tcoeff[n,m]$ which  differ for some particular $ m=m_0$ at $n=n_0$  but coincides for all lower-point correlators:
\be
\dcoeff[n,m] = 0, ~~\forall n < n_0; \qquad \dcoeff[n_0,m]  = 0, ~~\forall m \neq m_0~,\qq \d\cG_{n_0,m_0} \neq 0~.
\ee
The Ward identities imply that the higher-point functionals cannot coincide. However, all of these come with a higher power of $\kappa$. Therefore we can consider the state
\be
{1 \over \kappa^{n_0}} \left. {\partial \Psi_{\lambda}[g, \chi] \over \partial \lambda} \right|_{\lambda = 0} = \left(\dcoeff[n_0,m_0] + \Or[\kappa] \right) \Psi[g, \chi] ~,
\ee
which clearly has a good limit as $\kappa \to 0$. The notable feature above is that all the higher-order terms in \eqref{psiderstate} have disappeared.

It is useful to recast this in slightly different notation. Representing the Euclidean vacuum by $|0 \rangle$, and choosing the wavefunctional $\Psi[g,\chi]$ to correspond to this state, we see that the set of states that satisfy the constraints in the $\kappa \rightarrow 0$ limit can be written as
\be
\label{freestates}
|\Psi_{\text{ng}} \rangle = \int d \vec{y} d \vec{z} \,\delta G_{n_0,m_0}^{\vI\vJ}(\vec{y}, \vec{z})  h_{i_1 j_1}(y_1) \ldots h_{i_{n_0} j_{n_0}}(y_{n_0}) \chi(z_1) \ldots \chi(z_{m_0}) |0 \rangle~.
\ee
Our conclusion can be summarized in the following two points.
\begin{enumerate}
\item
The set of valid states in the nongravitational limit can be obtained by studying gravitational and matter fluctuations at late times, integrating them with a conformally invariant function on the late-time slice and acting on the Euclidean vacuum. This function is the difference of any two functions that obey the Ward identities of section \ref{subsecexpzg}. So it also obeys the Ward identities but without an inhomogeneous term.
\item
The smearing function in \eqref{freestates} is not arbitrary and is constrained by conformal invariance. This means that, even in the nongravitational limit, the effect of the constraints does not trivialize. This is consistent with the idea that when one takes the zero-coupling limit of a gauge theory, it is still necessary to impose the Gauss law on the Hilbert space.
\end{enumerate}

\paragraph{\bf Examples of states in the nongravitational limit:}
We now provide a few examples to help elucidate the idea above. To lighten the notation, we provide examples of states obtained by the action of matter-sector operators. It is simple to generalize this to consider states of gravitons, which exist for $d > 2$.

Conformal symmetry sharply constrains the Hilbert space at small ``particle number.''  Here, by particle number, we refer to the number of fields that must act on the Euclidean vacuum to produce the state.  There is a unique two-particle state with gravitons or with matter excitations corresponding to the fact that the two-point coefficient function is fixed up to an overall constant. Similarly, there is a unique three-particle state with scalar excitations and two possible three-particle states with graviton excitations. There exist an infinite number of four-particle states parameterized by  functions of conformally invariant cross ratios.

\begin{enumerate}
\item
{\bf Two-particle states.}
The unique two-particle matter state has the form
\be
|\chi \chi \rangle = \int d^d x_1 d^d x_2 {1 \over |x_1 - x_2|^{2 (d- \Delta)}} \chi(x_1) \chi(x_2) |0\rangle~.
\ee
Similarly, for $d > 2$, one can construct nontrivial two-particle states of free gravitons.
\item
{\bf Three-particle states.}
The three-point function of scalar operators is also fixed uniquely by conformal invariance up to an overall normalization. Therefore, we find the unique three-particle state
\be
|\chi \chi \chi \rangle = \int d^d x_1 d^d x_2 d^d x_3 {1 \over |x_1 - x_2|^{d-\Delta} |x_2 - x_3|^{d-\Delta} |x_1 - x_3|^{d-\Delta} } \chi(x_1) \chi(x_2) \chi(x_3) |0\rangle~.
\ee
\item
{\bf Four-particle states.}
Four-point functions that satisfy the Ward identities are undetermined up to a function of the conformal cross ratios. There is an infinite number of four-particle states that can be written in the form
\be
|\chi \chi \chi \chi \rangle =  \int d^d x_1 \dots d^d x_4 \,Q\!\le({x_{12} x_{3 4} \over x_{1 3} x_{2 4}}, {x_{1 2} x_{3 4} \over x_{2 3} x_{1 4}}\ri) \prod_{i < j} |x_{i j}|^{-{2(d- \Delta) \over 3}}   \chi(x_1) \chi(x_2) \chi(x_3) \chi(x_4) |0\rangle ~,
\ee
where $Q$ is an arbitrary function.
\end{enumerate}
Apart from graviton excitations, it is also possible to consider states with both graviton and matter excitations. These can be constructed using a procedure similar to the one above. 
There is no state with one graviton and two matter particles. This is because, as noted above, the corresponding correlation function, including its normalization, is completely fixed by the Ward identities and the two-point matter correlation function \cite{Osborn:1993cr}.

\subsection{Correspondence with Higuchi's construction}
We now show that the $\kappa \to 0$ limit of our construction described above corresponds precisely to Higuchi's construction of the Fock space for weakly-coupled gravity in de Sitter space. For simplicity, we discuss Higuchi's construction for a scalar field.   

\subsubsection{Review of Higuchi's proposal}
So far, we have been careful to discuss the metric and fields only on a single Cauchy slice. To make contact with Higuchi's construction we will briefly discuss the properties of fields in {\em spacetime}.

Consider a quantum field theory with a scalar field of mass $m$, $\chi$, propagating in a background de Sitter geometry with spacetime metric
\be
\label{globaldsmetric}
ds^2 = -d t^2 +  \r{cosh}^2 t\,d\Om_d^2; \qq d\Om^2_d =  {4 dx^2 \over (1+ |x|^2)^2}.
\ee

Then $\chi$ can be expanded in terms of solutions to the equations of motion that, at late times, have the asymptotic behaviour \cite{Chernikov:1968zm,Bros:1995js}
\be\label{chiasymp}
{\chi}^{\text{phys}}(t, x) \underset{t \rightarrow \infty}{\longrightarrow} e^{-\Delta t} \left({1 + |x|^2 \over 2} \right)^{\Delta} \chi(x) + e^{-\bar\Delta t} \left({1 + |x|^2 \over 2} \right)^{\bar{\Delta}} \bar\chi(x) ~,
\ee
where $\Delta$ and $\bar\D$ are the two solutions of the equation \eqref{massFormula}. We are interested in the operator
\be
\label{chilatetime}
\chi(x) = \lim_{t \rightarrow \infty} e^{\Delta t}  \left({1 + |x|^2 \over 2} \right)^{-\Delta} \chi^{\text{phys}}(t,x) ~,
\ee
which is well defined even when $\Delta$ has an imaginary part since even in that case the expression $e^{(\Delta - \bar\Delta) t}  \left({1 + |x|^2 \over 2} \right)^{\bar{\Delta}-\Delta} \bar{\chi}(x)$  can be neglected by the Riemann-Lebesgue lemma.  These rescaled late-time operators are precisely the ones that we have been studying in the previous sections. This can be seen by comparing \eqref{chilatetime} with \eqref{matterphysrescaled}.

Starting with the Euclidean vacuum we see that states of the form
\be
\label{samplestate}
|\Psi_{\text{seed}} \rangle =  \int d \vec{x} \, \qstate(\vec{x}) \chi(x_1) \ldots \chi(x_n) | 0 \rangle
\ee
span the Hilbert space in a nongravitational QFT where $\qstate$ is a square-integrable smearing function. More details and an oscillator construction can be found in \cite{Bros:1995js,Marolf:2008it,Marolf:2008hg}.

When the theory is coupled to gravity, it is necessary to impose the gravitational Gauss law even in the limit of arbitrarily weak coupling. The Gauss law requires that states be invariant under the de Sitter-isometry group $\confgrp$ \cite{Higuchi:1991tk,1975JMP....16..493M,1976JMP....17.1893M}.  This constraint can also be derived by integrating the Hamiltonian constraint \eqref{hamexplicit} with the Killing vectors of dS. But, except for the vacuum, no state of the form \eqref{samplestate} satisfies this constraint. 

Higuchi's proposal \cite{Higuchi:1991tm} was to consider the space of states obtained by ``averaging'' such seed states over the de Sitter isometry group
\be
|\Psi \rangle = \int d U U |\Psi_{\text{seed}} \rangle ~,
\ee
where $U$ is the unitary operator that implements the action of the de Sitter isometries in the quantum field theory and $d U$ is the Haar measure on this unitary group.  By construction we now have
\be
U |\Psi \rangle = |\Psi \rangle
\ee
for the action of any unitary element of the symmetry group.

The states $|\Psi \rangle$ are not normalizable in the original Hilbert space but Higuchi proposed a modified norm
\be
(\Psi,\Psi) = {1 \over \text{vol}(\confgrp)}\langle \Psi | \Psi \rangle = \int d U\langle \Psi_{\text{seed}}| U |\Psi_{\text{seed}} \rangle.
\ee
In \cite{Chandrasekaran:2022cip}, it was shown that this procedure can be understood in terms of imposing the equivalence relation $|\Psi_{\text{seed}} \rangle \sim U |\Psi_{\text{seed}} \rangle$ on the original Hilbert space. The final Hilbert space of equivalence classes is the same as the Hilbert space obtained by Higuchi's construction.

\subsubsection{Invariance of states under $\confgrp$}
We now show that the states \eqref{freestates} that we have found in the nongravitational limit  are invariant under the de Sitter isometries. 
To simplify the notation, we restrict to scalar states of the form
\be
\label{freescalarstates}
|\Psi \rangle = \int d \vec{x}  \, \delta G_{0,m}(x_1,\dots,x_m)  \chi(x_1) \ldots \chi(x_m) |0 \rangle~.
\ee
The inclusion of graviton states is simple but just requires us to keep track of some additional rotation matrices below.

The de Sitter isometries map the late-time boundary back to itself and act as conformal Killing vectors on it. (See \cite{Tod:2015hma} for a pedagogical explanation.) Their finite action at late times can be read off from \eqref{globaldsmetric} and is
\begin{align}\nt
{\text{translations}}: \quad \tilde{x}^{i} & =  x^{i} +  c^{i}, & \tilde{t} & =  t + \log{(1 + |\tilde{x}|^2) \over  (1 + |x|^2)} ; \\
{\text{rotations}}:  \quad\tilde{x}^{i} & =   R^{i}_{j}x^{j},& \tilde{t} & =  t; \\\nt
{\text{dilatations}}: \quad  \tilde{x}^{i} & =  \lambda x^{i}, & \tilde{t} & =  t + \log{(1 + |\tilde{x}|^2) \over \lambda (1 + |x|^2)}; \\\nt
{\text{SCTs}}: \quad \tilde{x}^i &= {x^i - \b^i |x|^2 \over 1 - 2 (\b\cdot x) + |\b|^2 |x|^2}  , & \tilde{t} &  =  t + \log{(1 + |\tilde{x}|^2) \over (1 + |x|^2)(1 - 2 (\b \cdot x )+ |\b|^2 |x|^2)},
\end{align}
where $c^i$ and $\b^j$ are  constant vectors, $R^{i}_{j}$ is a constant rotation matrix and $\lambda$ is a real number. Here we have neglected terms that vanish exponentially in $t$ since such terms are unimportant on the late-time boundary.  We note that the transformations above satisfy
\be
e^{\tilde{t} - t} =  \Lambda(x) \,{1 + |\tilde{x}|^2 \over 1 + |x|^2} ~,
\ee
where 
\be
\Lambda(x) = \le|\r{det}\Big( {\p \tilde{x}^i \/ \p x^j} \Big)\ri|^{-1/d}~.
\ee
Using \eqref{chilatetime},  we see that the unitary operator that implements this transformation on the fields acts as
\be
U \chi(\tilde{x}) U^{\dagger} = \Lambda(x)^{\Delta} \chi(x)~.
\ee
Therefore, using the transformation of $\delta G$ under conformal transformations as given in \eqref{finiteConf}, the state above transforms as
\be
\begin{split}
U |\Psi \rangle &= \int  d^d \tilde{x}_1\dots d^d \tilde{x}_m\, \delta G(\tilde{x}_1,\dots,\tilde{x}_m) U \chi(\tilde{x}_1) U^{\dagger} \ldots U \chi(\tilde{x}_m) U^{\dagger} |0 \rangle \\
& = \int  d^d \tilde{x}_1\dots d^d \tilde{x}_m\, \Big(\prod_{i=1}^m \Lambda(x_i)^{d-\Delta}\Big) \delta G(x_1,\dots,x_m) \big(\prod_{i=1}^m \Lambda(x_i)^{\Delta}\big) \chi(x_1) \ldots \chi(x_m) |0 \rangle \\ &= \int  d^d x_1\dots d^d x_m\, \delta G(x_1,\dots,x_m)  \chi(x_1) \ldots \chi(x_m) | 0 \rangle = |\Psi \rangle, 
\end{split}
\ee
and is therefore invariant. In the equalities above, we have used that $d^d\tilde{x}= \L(x)^{-d} d^dx $ which
ensures that the eventual expression has no factor of $\Lambda$.

\subsubsection{Lifting seed states \label{liftseed}}
It is also possible to obtain the seed states corresponding to \eqref{freestates}. The intuition is that the expression \eqref{freestates} has an implicit integral over the conformal group that can be pulled out to yield the seed state. This relies on the geometric observation that, in any number of dimensions, the conformal group can be used to fix three points 
leaving behind an unfixed $\sodminone$ that leaves those three points invariant. 

This can be made precise by considering the quantity
\be
f(x_1,x_2,x_3)=\int_{\confgrp} D\g\, \d^{(d)}(\hat{x}_1- \g x_1)\d^{(d)}(\hat{x}_2-\g x_2)\d^{(d)}(\hat{x}_3- \g x_3)  ~,
\ee
which must be invariant under an arbitrary conformal transformation $x_k\ra \g x_k$ by invariance of the Haar measure $D\g$. This is a conformally invariant function of three points and hence must be a constant, which can be normalized to $f(x_1,x_2,x_3)=\r{vol}(\sodminone)$. From this we obtain the identity
\bea\nt
1\= {1\/\r{vol}(\sodminone)} \int_{\confgrp} D\g \,\delta^{(d)}(x_1 - \g^{-1}\hat{x}_1)\delta^{(d)}(x_2 -\g^{-1} \hat{x}_2) \delta^{(d)}(x_3 - \g^{-1}\hat{x}_3)\\
&& \hspace{7cm}\times(\L(x_1) \L(x_2) \L(x_3))^{d},
\eea
using $\d^{(d)} (\tilde{x}-\g x) = \d^{(d)}(x- \g^{-1}\tilde{x}) \L(x)^{d}$.  

Inserting this in the expression \eqref{freescalarstates} removes the integrals over $x_1,x_2,x_3$ and gives
\be
\begin{split}
&|\Psi\rn = {1\/\r{vol}(\sodminone)} \int_{\confgrp} \,D\g \int d^d x_4\dots d^d x_m\,\d G_m(\g^{-1}\hat{x}_1,\g^{-1}\hat{x}_2,\g^{-1}\hat{x}_3,x_4,\dots,x_m) \\
& \hspace{2cm}\times (\L(\g^{-1}\hat{x}_1) \L(\g^{-1}\hat{x}_2) \L(\g^{-1}\hat{x}_3))^{d} \chi(\g^{-1}\hat{x}_1)\chi(\g^{-1}\hat{x}_2)\chi(\g^{-1}\hat{x}_3)\chi(x_4)\dots\chi(x_m)|0\rn \\
&= {1\/\r{vol}(\sodminone)} \int_{\confgrp} \,D\g \int d^d x_4\dots d^d x_m\,\big(\prod_{i=4}^m \Lambda(x_i)^{-d}\big)  \d G_m(\hat{x}_1,\hat{x}_2,\hat{x}_3,\tilde{x}_4,\dots,\tilde{x}_m)\\
& \hspace{2cm}\times  U {\chi}(\hat{x}_1)U^{-1}U\chi(\hat{x}_2)U^{-1}U\chi(\hat{x}_3)U^{-1}U\chi(\tilde{x}_4)U^{-1}\dots U \chi(\tilde{x}_m)U^{-1}|0\rn \\
&= {1\/\r{vol}(\sodminone)} \int_{\confgrp} \,D\g \int d^d \tilde{x}_4\dots d^d \tilde{x}_m\,\d G_m(\hat{x}_1,\hat{x}_2,\hat{x}_3,\tilde{x}_4,\dots,\tilde{x}_m)\\
& \hspace{6cm}\times  U \chi(\hat{x}_1)\chi(\hat{x}_2)\chi(\hat{x}_3)\chi(\tilde{x}_4)\dots\chi(\tilde{x}_m)|0\rn~,
\end{split}
\ee
where in the second step we have applied $\g$ to all the variables and used the conformal transformation properties of $\d G_m$ and $\chi$ to simplify the answer. 

We recognize that this takes the form of a group average
\be
|\Psi\rn =\int DU \, U |\Psi_\r{seed}\rn~ ~,
\ee
where the seed state is given by
\bea
|\Psi_\r{seed}\rn\= {1\/\r{vol}(\sodminone)}\int d^d x_4\dots d^d x_m\d G_m(\hat{x}_1,\hat{x}_2,\hat{x}_3,x_4,\dots,x_m)\-
&& \hspace{6cm}\times\chi(\hat{x}_1)\chi(\hat{x}_2)\chi(\hat{x}_3)\chi(x_4)\dots \chi(x_m)|0\rn~.
\eea
As a result, we obtain a valid seed state  by simply fixing three points to arbitrary positions $\hat{x}_1,\hat{x}_2,\hat{x}_3$ and dividing by the volume of $\sodminone$.  A convenient choice is to take $\hat{x}_1= (0,\dots,0),\hat{x}_2=(1,0,\dots,0) ,\hat{x}_3=\infty$.

\section{Discussion}
In this paper, we studied solutions to the WDW equation in asymptotically de Sitter space. A natural clock in de Sitter space is provided by the volume of the Cauchy slices. We found that, in the limit of large volume, all solutions could be written as the product of a universal phase factor multiplied by a diffeomorphism-invariant functional with simple Weyl transformation properties. This result is derived in section \ref{secsol} and we argued that the structural form of the solution is valid at all orders in perturbation theory. The Euclidean vacuum  is well known to have these properties but the new result is that all states in the theory have these properties.

In section \ref{sechilbastheory}, we showed that a solution could be specified by providing a list of coefficient functions that obey the same constraints as correlation functions of a CFT. These functions are related to one another by Ward identities. A specification of these functions provides a complete description of the state but it can also be said to specify a ``theory''. In this sense, the space of solutions to the WDW equation is similar to theory space.

In section \ref{secpert}, we rewrote these solutions in a basis of ``excitations'' about the Euclidean vacuum. Here, each solution is written as a series of multilinear functionals of the metric and other fields multiplied with the wavefunctional for the Euclidean vacuum. These excitations must  again obey the constraints of conformal invariance and the Ward identities.  It is shown in section \ref{secpert} that, in the nongravitational limit, these states reduce to those constructed by Higuchi through ``group averaging''. Therefore, our procedure not only provides a systematic justification for Higuchi's result, it specifies how the result should be generalized away from zero gravitational coupling.

In our analysis, we  have assumed that the Cauchy slice has the topology of the sphere $S^d$. 
From a technical perspective, there does not appear to be any immediate obstruction to generalizing our asymptotic solution 
to alternate topologies but some of the interesting physics might require us to go beyond perturbation theory. It would be interesting to examine the effects of change in topology 
 and to  understand whether a nonperturbative analysis reveals additional restrictions on the Hilbert space.

In an accompanying paper \cite{dssecond2023}, we describe a norm on the space of solutions to the WDW equation. The norm we propose is to simply average the
square of the absolute value of the wavefunctional   over the space of all possible metrics and matter fluctuations.  In \cite{dssecond2023}, we show that the states examined in section \ref{secpert} yield a normalizable basis for the Hilbert space. We also show that this prescription for the norm reduces, in the nongravitational limit, to the group-averaged norm proposed by Higuchi but differs at finite $\kappa$.  

In \cite{dssecond2023}, we define and study ``cosmological correlators''  in a gravitational theory. We find that these correlators display a remarkable property: knowledge of cosmological correlators in an arbitrary small region of the late-time spatial slice suffices to fix them everywhere, even in an arbitrary state. This result relies on the observation that all states, and not just the Euclidean vacuum, are covariant under scale transformations and translations. This provides a  generalization of the principle of ``holography of information'' --- previously explored in AdS and in flat space \cite{Laddha:2020kvp,Chowdhury:2020hse,Raju:2020smc,Chowdhury:2021nxw,Raju:2021lwh,Chakravarty:2023cll,deMelloKoch:2022sul} --- to asymptotically de Sitter space.

One interesting implication of our analysis is that all states in the Hilbert space share the symmetries of the Hartle-Hawking state. The inflationary era was presumably described by a state from this Hilbert space. On the one hand, this strengthens arguments like \cite{Mata:2012bx} that are based only on  symmetries. But it makes the effort to extract early-universe physics from inflationary correlators  \cite{Baumann:2022jpr} more interesting since one must contend not only with inflationary physics but also the possible states of the system.

\section*{Acknowledgments}
We are grateful to Simon Caron-Huot, Abhijit Gadde, Rifath Khan, Alex Maloney, Harkirat Singh Sahota, Ashoke Sen and Sandip Trivedi for helpful discussions. We also acknowledge several discussions with the string theory group at ICTS-TIFR. S.R. would like to acknowledge the hospitality of the 12th Joburg Workshop on string theory, the Abu Dhabi meeting in theoretical physics and the workshop on observables in quantum gravity (IISER Mohali) where preliminary versions of these results were presented. S.R. is partially supported by a Swarnajayanti fellowship, DST/SJF/PSA-02/2016-17, of the Department of Science and Technology. J.C. is supported by the Simons Collaboration on Nonperturbative Bootstrap. Research at ICTS-TIFR is supported by the Department of Atomic Energy, Government of India, under Project Identification Nos. RTI4001.
\section*{Appendix}
\appendix
\changelocaltocdepth{1}

\section{Wheeler-DeWitt expansion}\label{app:details}

In this Appendix we include technical details about the asymptotic expansion of the WDW equation and its solutions described in section \ref{secsol}.

\subsection{Rewriting the constraints \label{subapprewriteconst}}

We first explain how to rewrite the constraints in terms of intermediate variables to make the asymptotic expansion manifest. The original Hamiltonian constraint is
\bea
\label{appham}
\cH \= {2\k^2\/g}\le( \pi_{ij}\pi^{ij} - {1\/d-1}\pi^2\ri) -{1\/2\k^2} (R-2\L) +\cH_\r{matter} + \cH_{\text{int}} ~.
\eea
We define new variables $\Om, \gamma_{i j}$ and $\chi$ by the relations
\be
g_{ij}  =\Om^2 \g_{ij},\qq \chi  = \Om^{-\D}\cO,\qq \r{det}\,\g_{ij}=1~ ~,
\ee
and they can be written in terms of the original variables as
\be
\Om = \r{det}(g)^{1/2d},\qq \g_{ij} = {1\/\r{det}(g)^{1/d}}g_{ij},\qq O = \r{det}(g)^{\D/2d}\chi~.
\ee
Their variations are then given by
\bea
\d \Om\={1\/2d}\Om g^{ij}\d g_{ij}\\
\d \g_{ij} \= \Om^{-2}\le( \d g_{ij} - {1\/d}  g_{ij}g^{k\l}\d g_{k\l}\ri)\\
\d O  \= {1\/2d}\D O  g^{ij}\d g_{ij}+\Om^\D\d\chi~.
\eea
From the identification
\bea
\d\Psi \=  {\d\Psi\/\d g_{ij}}\d g_{ij}+{\d\Psi\/\d \chi}\d\chi = {\d\Psi\/\d\Om}\d\Om +{\d\Psi\/\d \g_{ij}}\d\g_{ij}+ {\d\Psi\/\d O }\d O ~,
\eea
we obtain the differential operators
\bea
\label{firstorderdiffrewr}
i\pi^{ij}\={\d\/\d g_{ij}}= {1\/2d} g^{ij} \le( \Om {\d\/\d\Om} +\D O{\d\/\d O}\ri)+\Om^{-2}\le(  {\d\/\d \g_{ij}}- {1\/d}  \g^{ij} \g_{k\l} {\d\/\d \g_{k\l}}\ri) ~,\\
i\pi \= g_{ij}{\d\/\d g_{ij} }={1\/2}\Om{\d\/\d\Om} + {1\/2} \D O  {\d\/\d O }~,\\
i\pi_\chi\={\d\/\d \chi}= \Om^\D{\d\/\d O }~.
\eea
We see that what appears is the traceless differential
\be
\dgtr^{ij} \equiv \Om^{-2}\le(  {\d\/\d \g_{ij}}- {1\/d}  \g^{ij} \g_{k\l} {\d\/\d \g_{k\l}}\ri)~.
\ee
A useful fact is that it can be written in terms of the original metric $g$ as
\be\label{deltahatmetric}
\dgtr^{ij} = {\d\/\d g_{ij}} -{1\/d}g^{ij}g_{k\l}{\d\/\d g_{k\l}}~.
\ee
The momentum then takes the form
\be
i \pi^{ij} = {1\/2d}g^{ij}\le( \Om {\d\/\d\Om} +\D O{\d\/\d O}\ri) +\dgtr^{ij}~,
\ee
and so the kinetic piece is
\be
\pi_{ij}\pi^{ij} -{1\/d-1}\pi^2 = {1\/4d(d-1)} \le( \Om {\d\/\d\Om} +\D O{\d\/\d O}\ri)^2-g_{ik} g_{j\l} \dgtr^{ij}\dgtr^{k\l} ~,
\ee
using the tracelessness condition $g_{ij}\dgtr^{ij}=0$ to cancel off-diagonal terms.

Now note that the Hamiltonian constraint, \eqref{appham}, involves a composition of two such differential operators. This yields terms where the second differential operator acts on the variable coefficients that appear in \eqref{firstorderdiffrewr} and produces the divergent expression, $\delta(0)$. Such terms can also arise if the second-order functional derivative acts on a local expression.  We discuss these terms further in Appendix \ref{app:contact} but we drop these terms for now. As explained in subsection \ref{subsecsolalg}, this issue does not affect our leading-order analysis.

The Hamiltonian constraint is then
\be
\cH ={2\k^2\/\Om^{2d}}\le[  {1\/4d(d-1)}\le(\Om{\d\/\d\Om} + \D O{\d\/\d O}\ri)^2-g_{ik} g_{j\l} \dgtr^{ij}\dgtr^{k\l}\ri]-{1\/2\k^2} (R-2\L) +\cH_\r{matter} + \cH_{\text{int}}
\ee
and for a scalar field we have
\be
\cH_\r{matter} = -{1\/2}g^{-1}\le( {\d\/\d \chi}\ri)^2 +{1\/2}(g^{ij}\p_i\chi\p_j\chi+ m^2\chi^2)~.
\ee
We obtain the form given in  \eqref{Hpsi} after rewriting the second term in the bracket  in terms of $\g_{ij}$ using that
\bea\nt
g_{ik}g_{j\l}\dgtr^{ij}\dgtr^{k\l}\= \g_{ik}\g_{j\l}\le( {\d\/\d \g_{ij}}- {1\/d}   \g^{ij} \g_{ab} {\d\/\d \g_{ab}} \ri)\le( {\d\/\d \g_{k\l}}- {1\/d} \g^{k\l} \g_{cd}  {\d\/\d \g_{cd}} \ri)\\\label{deltagammaidentity}
\=\le(\g_{ik}\g_{j\l}-{1\/d}\g_{ij}\g_{k\l}\ri){\d\/\d \g_{ij}}{\d\/\d \g_{k\l}}~.
\eea

\subsection{Normal ordering prescription}\label{app:contact}

We describe a natural choice of normal ordering prescription that gets rid of the $\delta(0)$ terms appearing at leading order when acting with the Hamiltonian constraint on an asymptotic wavefunctional
\be
\Psi = e^{i\cF},\qq \cF = \int d^d x \sqrt{g}\le(- {(d-1)\/\k^2}   +b_\b \chi^2 +\dots\ri)  ~,
\ee
where $\dots$ corresponds to subleading pieces in our asymptotic expansion.

The leading contributions come from the derivatives with respect to the Weyl factor. We can choose the normal ordering prescription
\bea
:\cH: \=  {2\k^2 \/4d(d-1)} {1\/\Om^d}\le(\conf {\delta \over \delta \conf}  + \D :O {\d\/\d O}:  \ri){1\/\Om^d}\le(\conf {\delta \over \delta \conf}  + \D :O {\d\/\d O}:  \ri) \- 
&&+ {\L\/\k^2}  -{1\/2} \conf^{2 (\Delta-d)} {\delta^2 \over \delta O^2}+{1\/2} m^2 \conf^{-2 \D} O ^2 +\cH_\r{sub} ~,
\eea
where $\cH_\r{sub}$ corresponds to terms that are subleading when acting on $\Psi$. For the matter we choose the normal ordering
\be
: O{\d\/\d O}:  ={1\/2}\le( O {\d\/\d O} + {\d\/\d O} O \ri) = O {\d\/\d O} +{1\/2}\d(0)~.
\ee
Recalling that $\sqrt{g}=\conf^d$, it is clear that the choice of ordering  cancels the $\d(0)$ appearing in the leading gravity piece. At leading order in the matter sector, we have
\bea
{:\cH: \Psi\/\Psi} \= {2\k^2\/4d(d-1)} {2\/\Om^d}\le(i \Om {\d X_d\/\d\Om} \ri) {\D\/2 }\d(0) - ib_\b \Om^{-d}\d(0) \-
\=- {i\/\Om^d} {\D\/2 }\d(0) - {ib_\b\/\Om^d}\d(0)~,
\eea
which vanishes since $b_\b = -{\D\/2}$.

The contribution of second order derivatives on subleading terms in the gravitational and matter part of the solution produces terms that compete with the remainder term in \eqref{solansatz}. Therefore, these terms are important for understanding finite-time physics but not for the asymptotic form of the solution. This is to be expected since finite-time physics should depend on the details of the UV-completion whereas the form of the Hilbert space can be determined more easily.

\subsection{Anomaly in $d=4$}\label{app:higherorder}

For pure Einstein gravity, the term $X_{d-4}$ satisfies the equation
\be
\Om{\d X_{d-4} \/\d\Om} =-{2\k^2\/\Om^d}\le( g_{ik}g_{j\l} (\dgtr^{ij} X_{d-2})(\dgtr^{k\l} X_{d-2})-{1\/4d(d-1)}\le( \Om {\d X_{d-2}\/\d\Om}\ri)^2\ri)~.
\ee
Using that $\d_g^{ij}$ it the traceless part of the variation with respect to $g_{ij}$, we see that
\be
\d_g^{ij}X_{d-2} = -{1\/2(d-2)\k^2}\sqrt{g}\le(R^{ij}-{1\/d} g^{ij}R\ri)~ ~,
\ee
so that we have
\be
g_{ik}g_{j\l} (\dgtr^{ij} X_{d-2})(\dgtr^{k\l} X_{d-2})= {\Om^{2d}\/4(d-2)^2\k^4} \le( R_{ij}R^{ij}-{1\/d}R^2\ri)~.
\ee
The second contribution takes the form
\be
{1\/4d(d-1)}\le( \Om {\d X_{d-2}\/\d\Om}\ri)^2 = {1\/16d(d-1)\k^4} \Om^{2d}R^2~,
\ee
and so the equation becomes
\be
\Om{\d X_{d-4} \/\d\Om} =-{1\/2(d-2)^2\k^2}\sqrt{g}\le( R_{ij}R^{ij}-{d\/4(d-1)}R^2\ri)~.
\ee
In $d\neq 4$, this equation can be integrated to give
\be
X_{d-4} = -{1\/2(d-2)^2(d-4)\k^2}\int d^dx\,\sqrt{g}\le( R_{ij}R^{ij}-{d\/4(d-1)}R^2\ri) ~,
\ee
which matches with the holographic renormalization results, see \eg (B.4) in \cite{deHaro:2000vlm}.

In $d=4$, we obtain the equation
\be
\Om {\d X_0\/\d\Om} =- {1\/8\k^2} \sqrt{g}\le( R_{ij}R^{ij}-{1\/3}R^2\ri) ~,
\ee
which leads to 
\be
\Om {\d\/\d\Om} e^{i X_0} = \cA_4 e^{i X_0},\qq \cA_4 \equiv  -{i\/8\k^2} \sqrt{g}\le( R_{ij}R^{ij}-{1\/3}R^2\ri)~.
\ee
We recognize the trace anomaly equation for the CFT partition function $Z = e^{i X_0}$. The anomaly can be written as
\be
\cA_4  ={1\/16\pi^2}\sqrt{g}(-a E_4+c W_{abcd}W^{abcd})
\ee
using the Euler density and Weyl squared curvature
\be
\begin{split}\label{EulerWeyl}
E_4 & = R_{abcd}R^{abcd} -4 R_{ab}R^{ab} + R^2~,\\
W_{abcd}W^{abcd} & = R_{abcd}R^{abcd} - 2 R_{ab}R^{ab}+{1\/3}R^2~,
\end{split}
\ee
with the anomaly coefficients
\be
a=c= -{i \pi^2\/ \k^2} = -{i\pi\/8G_N}~.
\ee
This is, up to the factor of $-i$, the anomaly of a holographic CFT$_4$ obtained using holographic renormalization in AdS$_5$ \cite{Balasubramanian:1999re}.

\subsection{Subleading matter term}\label{app:subleadmatter}
The subleading matter term $Y_{\beta-2}$ is determined by the equation
\begin{equation}
	\frac{2}{\sqrt{g}}\enc{g_{ij}\frac{\delta}{\delta g_{ij}} + b_\beta \chi\frac{\delta}{\delta \chi}} Y_{\beta-2} - \frac{b_\beta }{2(d-1)}R \chi^2 +\frac{1}{2}g^{ij}\partial_i\chi \partial_j \chi=0.
\end{equation}
An ansatz for the solution is to take the local and diffeomorphism invariant functional
\begin{equation}
	Y_{\beta-2}= c_1 \underbrace{\int d^dx\, \sqrt{g} R\chi^2}_{\text{I}} + c_2 \underbrace{\int d^dx\, \sqrt{g} g^{ij} \partial_i \chi \partial_j \chi}_{\text{II}},
\end{equation}
where $c_{1,2}$ are undetermined coefficients. These functionals I and II have been chosen due to their $\conf^{\beta-2}=\conf^{d-2\Delta-2}$ scaling. Defining
\begin{equation}
	\delta_1 = \frac{g_{ij}}{\sqrt{g}} \frac{\delta}{\delta g_{ij}},\qquad \delta_2=\frac{\chi}{\sqrt{g}} \frac{\delta}{\delta\chi}~,
\end{equation}
we have the formulae
\begin{align}
	\delta_1 \text{I}&=\frac{g_{ij}}{\sqrt{g}} \frac{\delta}{\delta g_{ij}} \int d^dx\, \sqrt{g} R\chi^2 = \enc{\frac{d}{2}-1} R\chi^2 - 2(d-1) \enc{(\nabla\chi)^2 + \chi\square\chi},\\
	\delta_1 \text{II}&=\frac{g_{ij}}{\sqrt{g}} \frac{\delta}{\delta g_{ij}} \int d^dx\, \sqrt{g} g^{ij} \partial_i \chi \partial_j \chi = \enc{\frac{d}{2}-1} (\nabla \chi)^2,\\
	\delta_2\text{I}&=\frac{\chi}{\sqrt{g}} \frac{\delta}{\delta\chi}\int d^dx\, \sqrt{g} R\chi^2 = 2R\chi^2,\\
	\delta_2\text{II}&=\frac{\chi}{\sqrt{g}} \frac{\delta}{\delta\chi}  \int d^dx\, \sqrt{g} g^{ij} \partial_i \chi \partial_j \chi = -2\chi\square\chi~.
\end{align}
Our equation now becomes
\begin{equation}
	(\delta_1  + b_\beta\delta_2)(c_1\text{I}+ c_2\text{II} )= -\frac{1}{4} (\nabla\chi)^2 + \frac{b_\beta}{4(d-1)} R\chi^2~.
\end{equation}
Requiring that the $\chi\square\chi$ term cancels out from the left side gives
\begin{equation}
	-2(d-1)c_1 -2b_\beta c_2=0 \quad \implies\quad c_2 = -\frac{(d-1)}{b_\beta} c_1~.
\end{equation}
Matching the coefficients of $R\chi^2$ on both sides then gives
\begin{equation}
	c_1\enc{\frac{d-2}{2}+2b_\beta} = \frac{b_\beta}{4(d-1)}~.
\end{equation}
Solution to above equation gives $c_1$ and the proportionality gives $c_2$ as
\begin{equation}
	c_1 = \frac{b_\beta}{2(d-1)(d-2+4b_\beta)}, \qquad c_2 = -\frac{1}{2(d-2+4b_\beta)}.
\end{equation}
We can see that this choice also matches the coefficient of $(\nabla\chi)^2$ on both sides, meaning the system of equations was overdetermined, albeit with a solution. Substituting $b_\beta = -\Delta/2$ from \eqref{eq:b_equation} we have,
\begin{equation}
		Y_{\beta-2} = -\frac{1}{2(d-2-2\Delta)} \int d^dx \, \sqrt{g} \enc{g^{ij}\partial_i\chi\partial_j\chi+\frac{\Delta}{2(d-1)}R\chi^2}.
\end{equation}

\section{Derivation of the Ward identities}\label{app:Ward}

In this Appendix, we derive the Ward identities for the coefficient functions. We have
\be
Z[g,\chi] = \exp\left[\sum_{m,n} \kappa^n \cG_{n,m}[h, \ldots h, \chi, \ldots \chi] \right] ~,
\ee
where we remind the reader of the definition of the multi-linear functionals, $\coeff[n,m]$  that take $n$ tensor fields and $m$ scalar fields as input and return a $c$-number. They are defined as
\bea
&&\coeff[n,m][h^{(1)}, \ldots h^{(n)}, \chi^{(1)}, \ldots \chi^{(m)}] \\\nt
&\equiv& {1 \over n! m!} \int  d^d y_1\dots d^d y_n d^d z_1\dots d^d z_m\,G_{n,m}^{\vec{i} \vec{j}}(\vec{y}, \vec{z})   h^{(1)}_{i_1 j_1}(y_1) \ldots h^{(n)}_{i_n j_n}(y_n) \chi^{(1)}(z_1) \ldots \chi^{(m)}(z_m) ~,
\eea
so that
\be
 G_{n,m}^{\vec{i} \vec{j}}(\vec{y}, \vec{z}) = {\d^n\/\d h_{i_1j_1}(y_1)\dots \d h_{i_n j_n}(y_n)} {\d^m\/\d \chi(z_1)\dots \chi(z_m)}\coeff[n,m][h,\dots,h,\chi,\dots,\chi]~.
\ee
Under a diffeomorphism and Weyl transformation, we have
 \be
\d_{(\xi,\vphi)} g_{i j}  =\n_i\xi_j+\n_j\xi_i +2 \vphi g_{i j},\qq \d_{(\xi,\vphi)}\chi = \xi^i\p_i\chi -\D \vphi\chi
\ee
so that  $h_{ij}$ transforms as
\be
\label{htransformapp}
\kappa \delta h_{ij} = \kappa H_{ij} + I_{ij} ~,
\ee
where 
\be
H_{ij} = \cL_\xi h_{ij}+2\vphi h_{ij},\qq I_{ij} = \p_i\xi^k \delta_{j k} +\p_j\xi^k \delta_{i k} + 2\vphi\d_{ij} ~,
\ee
and the definition of the Lie derivative gives
\be
\cL_\xi h_{ij} = \xi^k\p_k h_{ij} + \p_i \xi^k h_{kj}+ \p_j \xi^k h_{ik}~.
\ee
We then have the variation 
\be
\begin{split}
\delta_{(\xi,\vphi)} \log Z &= \sum_{n,m}  \kappa^{n}(\d_{(\xi,\vphi)} \log Z)_{n,m}~,
\end{split}
\ee
where we have collected terms according to the expansion in $\k$:
\bea
(\d_{(\xi,\vphi)} \log Z)_{n,m} \=  (n+1)\coeff[n+1,m][I, h, \ldots h, \chi, \ldots \chi] + n  \coeff[n,m][H, h, \ldots h, \chi, \ldots \chi] \-
 && + m  \coeff[n,m] [h, \ldots h, \delta \chi, \ldots \chi]~.
\eea
The Weyl and diffeomorphism invariance can be summarized by the identity
\be
\d_{(\xi,\vphi)} \log Z = \int d^d x\,\vphi(x)\cA_d(x)~.
\ee
That this is equivalent to the equations \eqref{Zidentities} can be proven by taking the functional derivatives with respect to $\vphi(x)$ and $\xi_k(x)$ . This follows from the fact that
\bea
{\d\/\d\vphi(x)} \d_{(\xi,\vphi)} \log Z \= \le(2 g_{ij}{\d\/\d g_{ij}} - \D \chi {\d\/\d\chi} \ri)\log Z~,\-
{\d\/\d\xi_j(x)} \d_{(\xi,\vphi)} \log Z \= \le(-2 \sqrt{g}  \nabla_{i} {1\/\sqrt{g} }{\delta \over \delta g_{i j}} +g^{i j} \p_i \chi {\delta \over \delta \chi} \ri)\log Z~.
\eea

\subsection{Trace identity}

The trace identity is obtained by considering only a Weyl transformation. Under this we have 
\be
H_{ij}=2\vphi h_{ij},\qq I_{ij}=2\vphi \d_{ij},\qq \d_\vphi \chi = -\D\vphi\chi~.
\ee
As a result, we get
\bea\nt
(\d_\vphi \log Z)_{n,m}\=2 (n+1)\cG_{n+1,m}[ \vphi \d ,h,\dots,h,\chi,\dots,\chi]+  2n \cG_{n,m}[\vphi h,h,\dots,h,\chi,\dots,\chi]  \\
&& -m \D \cG_{n,m}[h,\dots,h,\vphi\chi,\chi,\dots,\chi]~.
\eea
We can expand the anomaly in the $\k$ to get
\be
\cA_d(x) = \sum_{n}{\k^n\/n!} \int d^d y_1\dots d^d y_n\,\cA_d^{\vec{i}\vec{j}}(x,\vec{y}) h_{i_1j_1}(y_1)\dots h_{i_n j_n}(y_n)~.
\ee
This defines the coefficients $\cA_d^{\vec{i}\vec{j}}(x,\vec{y})$ which can be recovered as the functional derivative
\be
\k^n \cA_d^{\vec{i}\vec{j}}(x,\vec{y}) =  {\d^n\/\d h_{i_1j_1}(y_1)\dots \d h_{i_nj_n}(y_n)}\cA_d(x)~.
\ee
As $\cA_d(x)$ is local, these coefficients are ultralocal, in the sense that
\be
\cA_d^{\vec{i}\vec{j}}(x,\vec{y}) = A_d^{\vec{i}\vec{j}}(x)\prod_{a=1}^n \d^{(d)}(y_a-x) ~,
\ee
where $ A_d^{\vec{i}\vec{j}}(x)$ are the coefficients appearing in expanding $\cA_d(x)$ in the metric.

The Weyl transformation equation then becomes
\be
(\d_\vphi \log Z)_{n,m} = \d_{m,0} {1\/n!} \int d^d x d^d y_1\dots d^d y_n\,\vphi(x)\cA_d^{\vec{i}\vec{j}}(x,\vec{y}) h_{i_1j_1}(y_1)\dots h_{i_n j_n}(y_n)~.
\ee
To obtain the relations satisfied by the coefficient functions $G_{n,m}^{\vec{i}\vec{j}}(\vec{y},\vec{z})$, we take $n$ functional derivatives with respect to the metric and $m$ functional derivatives with respect to the matter. 

For the LHS, we get
\bea\nt
 && {\d^n\/\d h_{i_1j_1}(y_1)\dots \d h_{i_nj_n}(y_n)} {\d^m\/\d \chi(z_1)\dots \chi(z_m)}(\d_\vphi \log Z)_{n,m}= 2 \int d^d x\,\vphi(x) \d_{ij} G_{n+1,m}^{ij\vec{i}\vec{j}}(x,\vec{y},\vec{z})\\
&&\hspace{2cm}+ \le[ 2 (\vphi(y_1)+\dots+\vphi(y_n))-\D(\vphi(z_1)+\dots+\vphi(z_m)) \ri]G_{n,m}^{\vec{i}\vec{j}}(\vec{y},\vec{z}) ~.
\eea
The equality with the RHS implies that 
\be
{\d^n\/\d h_{i_1j_1}(y_1)\dots \d h_{i_nj_n}(y_n)} {\d^m\/\d \chi(z_1)\dots \chi(z_m)} (\d_\vphi \log Z)_{n,m} = \d_{m,0}\int d^dx \,\vphi(x) \cA_d^{\vec{i}\vec{j}}(x,\vec{y})~.
\ee
Equating these two expressions and taking the functional derivative with respect to $\vphi(x)$ gives the trace identity
\be
2 \d_{ij}G_{n+1,m}^{ij\vec{i}\vec{j}}(x,\vec{y},\vec{z}) = \le(-2 \sum_{a=1}^n \d^{(d)}(x-y_a)+\D \sum_{b=1}^m \d^{(d)}(x-z_b) \ri)G_{n,m}^{\vec{i}\vec{j}}(\vec{y},\vec{z}) +  \d_{m,0}\cA_d^{\vec{i}\vec{j}}(x,\vec{y})~.
\ee

\subsection{Divergence identity}

Under a diffeomorphism we have
\bea
H_{ij}\=\cL_\xi h_{ij}=  \xi^k\p_k h_{ij} + \p_i \xi^k h_{kj}+ \p_j \xi^k h_{ik} ~,\\ 
I_{ij}\=\cL_\xi \d_{ij}=\p_i\xi^k \d_{jk}+\p_j\xi^k\d_{ki} ~,\\
\d_\xi \chi \= \cL_\xi \chi = \xi^k\p_k \chi~.
\eea
The Ward identity is
\be
\d_\xi \log Z= 0 ~,
\ee
which we expand as 
\bea\nt
0= (\d_\xi \log Z)_{n,m} \= (n+1)\cG_{n+1,m}[ \cL_\xi \d ,h,\dots,h,\chi,\dots,\chi] + n \cG_{n,m}[ \cL_\xi h,h,\dots,h,\chi,\dots,\chi] \\
&& +m \cG_{n,m}[h,\dots,h, \cL_\xi \chi,\chi,\dots,\chi] ~.
\eea
As above we take functional derivatives and obtain
\bea
0 \= {\d^n\/\d h_{i_1j_1}(y_1)\dots \d h_{i_nj_n}(y_n)} {\d^m\/\d \chi(z_1)\dots \chi(z_m)}(\d_\xi \log Z)_{n,m}\-
 \= 2 \int d^d x\, \d_{jk}\p_i\xi^k(x) G_{n+1,m}^{ij \vec{i}\vec{j}}(x,\vec{y},\vec{z}) -\sum_{b=1}^m {\p\/\p z_b^k}(\xi^k(z_b) G_{n,m}^{\vec{i}\vec{j}}(\vec{y},\vec{z}))\-
 && + \sum_{a=1}^n \le[-{\p\/\p y_a^k} (\xi^k(y_a) G_{n,m}^{\vec{i}\vec{j}}(\vec{y},\vec{z}))+ ( \d_{i_a'}^{i_a}\p_{j_a'}\xi^{j_a}(y_a)+ \d_{j_a'}^{j_a}\p_{i_a'}\xi^{i_a}(y_a))G_{n,m}^{\vec{i}'\vec{j}'}(\vec{y},\vec{z})\ri] ~,
 \eea
 where in the bracketed expression, we use the notation
\be
\label{ipjpdef}
(\vec{i}'\vec{j}') = (i_1, j_1 \ldots i_{a-1}, j_{a-1}, i_{a}', j_{a}', i_{a+1}, j_{a+1} \ldots i_{n}, j_{n})
\ee
for the current $a$ in the sum. We now take the functional derivative with respect to $\xi^k(x)$ which can be done by simply replacing $\xi^{\l}(x')$ by $\d_{k}^{\l}\d^{(d)}(x-x')$. This leads to the divergence identity
\bea
&& 2  \d_{jk}\p_i G_{n+1,m}^{i j\vec{i}\vec{j}}(x,\vec{y},\vec{z})=-\sum_{b=1}^m {\p\/\p z_b^k} \Big[ \d^{(d)}(x-z_b)G_{n,m}^{\vec{i}\vec{j}}(\vec{y},\vec{z})\Big]\-
&& +\sum_{a=1}^n \Big[ -{\p\/\p y_a^k} \big[ \d^{(d)}(x-y_a) G_{n,m}^{\vec{i}\vec{j}}(\vec{y},\vec{z}) \big]+ G_{n,m}^{\vec{i}'\vec{j}'}(\vec{y},\vec{z})\big( \d_{i_a'}^{i_a}\d^{j_a}_k {\p\/\p y^{j_a'}}+ \d_{j_a'}^{j_a} \d^{i_a}_k {\p\/\p y^{i_a'}}\big)\d^{(d)}(x-y_a)\Big]~.
\eea

\subsection{Conformal symmetry}\label{app:Wardconf}

The conformal symmetry is obtained by combining a diffeomorphism $\xi$ and a Weyl transformation with $\vphi = -\p_k \xi^k/d$ so that
\be
I_{ij} = \p_i \xi^k \d_{jk}+ \p_j\xi^k \d_{ik} - {2\/d}\d_{ij}\p_k\xi^k = 0~.
\ee
This is  achieved by taking  $\xi$ to be a conformal Killing vector on the sphere. We then have
\be
\d h_{ij}=H_{ij}=\xi^k\p_k h_{ij} + \p_i \xi^k h_{kj}+ \p_j\xi^k h_{ik} -{2\/d}\p_k\xi^k h_{ij},\qq \d \chi = \xi^i\p_i\chi+{\D\/d} \p_k\xi^k \chi~.
\ee
The variation of $\log Z$ doesn't mix different terms in the expansion as $I=0$. So we get
\be
(\d \log Z)_{n,m} =  n  \coeff[n,m][H, h, \ldots h, \chi, \ldots \chi] + m  \coeff[n,m] [h, \ldots h, \delta \chi, \ldots \chi] ~,
\ee
and the constraint gives
\be
(\d \log Z)_{n,m}= - \d_{m,0} {1\/dn!} \int d^d x d^d y_1\dots d^d y_n\, \p_k \xi^k(x)\cA_d^{\vec{i}\vec{j}}(x,\vec{y}) h_{i_1j_1}(y_1)\dots h_{i_n j_n}(y_n)~.
\ee
Taking successive functional derivatives gives
\be
\resizebox{0.98\textwidth}{!}{$
\begin{aligned}
 &{\d^n\/\d h_{i_1j_1}(y_1)\dots \d h_{i_nj_n}(y_n)} {\d^m\/\d \chi(z_1)\dots \chi(z_m)} (\d \log Z)_{n,m} \\
&= \sum_{a=1}^n \Big[\big( \d_{i_a'}^{i_a}\p_{j_a'}\xi^{j_a}(y_a)+ \d_{j_a'}^{j_a}\p_{i_a'}\xi^{i_a}(y_a)\big)G_{n,m}^{\vec{i}'\vec{j}'}(\vec{y},\vec{z}) - {\p\/\p y_a^k} \big[\xi^k(y_a) G_{n,m}^{\vec{i}\vec{j}}(\vec{y},\vec{z})\big]  - {2\/d} \p_k\xi^k(y_a) G_{n,m}^{\vec{i}\vec{j}}(\vec{y},\vec{z}) \Big] \\
 &+\sum_{b=1}^m \Big[- {\p\/\p z_b^k}\big[\xi^k(z_b) G_{n,m}^{\vec{i}\vec{j}}(\vec{y},\vec{z})\big] +{\D\/d}\p_k\xi^k(z_b) G_{n,m}^{\vec{i}\vec{j}}(\vec{y},\vec{z})\Big]~,\\
 &= \sum_{a=1}^n \Big[ - \xi^k(y_a){\p\/\p y_a^k}  G_{n,m}^{\vec{i}\vec{j}}(\vec{y},\vec{z})- {2+d\/d} \p_k \xi^k(y_a)G_{n,m}^{\vec{i}\vec{j}}(\vec{y},\vec{z})+ \big( \d_{i_a'}^{i_a}\p_{j_a'}\xi^{j_a}(y_a)+ \d_{j_a'}^{j_a}\p_{i_a'}\xi^{i_a}(y_a)\big)G_{n,m}^{\vec{i}'\vec{j}'}(\vec{y},\vec{z})\Big] \\
 &+\sum_{b=1}^m \Big[-\xi^k(z_b)  {\p\/\p y_b^k}G_{n,m}^{\vec{i}\vec{j}}(\vec{y},\vec{z})+{\D-d\/d}\p_k\xi^k(z_b) G_{n,m}^{\vec{i}\vec{j}}(\vec{y},\vec{z})\Big]~.
\end{aligned}$}
\ee
Ignoring the anomaly which is ultralocal, we get the conformal Ward identity
\be
\label{app:confWard}
\resizebox{0.98\textwidth}{!}{$
\begin{aligned}
0 &=\sum_{a=1}^n \le[ - \xi^k(y_a){\p\/\p y_a^k}  G_{n,m}^{\vec{i}\vec{j}}(\vec{y},\vec{z})- {2+d\/d} \p_k \xi^k(y_a)G_{n,m}^{\vec{i}\vec{j}}(\vec{y},\vec{z})+ \le( \d_{i_a'}^{i_a}\p_{j_a'}\xi^{j_a}(y_a)+ \d_{j_a'}^{j_a}\p_{i_a'}\xi^{i_a}(y_a)\ri)G_{n,m}^{\vec{i}'\vec{j}'}(\vec{y},\vec{z})\ri] \\
 &+\sum_{b=1}^m \le[-\xi^k(z_b)  {\p\/\p z_b^k}G_{n,m}^{\vec{i}\vec{j}}(\vec{y},\vec{z})+{\D-d\/d}\p_k\xi^k(z_b) G_{n,m}^{\vec{i}\vec{j}}(\vec{y},\vec{z})\ri]~
\end{aligned}$}
\ee
Under a finite conformal transformation $x\ra x'$, the Jacobian is
\be
J^{i}_{i'}(x)={\p x'^{i}\/\p x^{i'}} ~ ~,
\ee
and the scale factor is defined as 
\be
\Lambda(x) = |\r{det} \,J(x)|^{-1/d} ~.
\ee
The rotation matrix is defined as 
\be
R^{i}_{i'}(x)= \Lambda(x) J^{i}_{i'}(x) ~,
\ee
so that it satisfies $\r{det}\,R =1$.  

The finite conformal transformation takes the form
\be\label{app:finiteConf}
G^{\vec{i}\vec{j}}_{n,m}(\vec{y}',\vec{z}')= \Big(\prod_{a=1}^n R_{i_a'}^{i_a}(y_a) R^{j_a}_{j_a'}(y_a) \Lambda(y_a)^d\Big)\Big(\prod_{b=1}^m \Lambda(z_b)^{d-\D}\Big) G_{n,m}^{\vec{i}'\vec{j}'}(\vec{y},\vec{z})~.
\ee
We can check that this is the integrated version of the conformal Ward identity by expanding infinitesimally. Under an infinitesimal conformal transformation, we have
\be
x'_a = x_a + \xi (x_a)+\dots,\qq J^{i}_{i'}(x)= \d^{i}_{i'}+\p_{i'}\xi^{i}(x)+\dots,\qq \Lambda(x)=1-{1\/d} \p_k \xi^k(x)+\dots ~,
\ee
so that we get
\bea
G^{\vec{i}\vec{j}}_{n,m}(\vec{y}',\vec{z}')\= \prod_{a=1}^n( \d_{i_a'}^{i_a}+ \p_{i_a'}\xi^{i_a}(y_a))(\d^{j_a}_{j_a'}+\p_{j_a'}\xi^{j_a}(y_a))\le(1- {2+d\/d}\p_k\xi^k(y_a)\ri)\-
&& \times \prod_{b=1}^m \le(1+{\D-d\/d}\p_k\xi^k(z_b)\ri) G_{n,m}^{\vec{i}'\vec{j}'}(\vec{y},\vec{z})~,
\eea
which reproduces at linear order the conformal Ward identity \eqref{app:confWard}. This shows that \eqref{app:finiteConf} is the finite conformal transformation properties of the coefficient functions.

We note that the coefficient functions have the same symmetries of a CFT correlator:
\be
G^{\vec{i}\vec{j}}_{n,m}(\vec{y},\vec{z}) \sim \ln T^{i_1 j_1}(y_1)\dots T^{i_n j_n}(y_n)\phi(z_1)\dots \phi(z_m)\rn_\r{CFT} ~,
\ee
where $\phi$ is an operator of dimension $d-\D$ and $T^{ij}$ is an operator of spin $2$ and dimension $d$.

\bibliography{references}

\end{document}